\documentclass[useAMS,usenatbib]{mn2e}

\usepackage{verbatim,graphicx,epsfig,dcolumn,color}
\definecolor{red}{rgb}{0.75,0.0,0.0}
\definecolor{green}{rgb}{0.0,0.75,0.0}
\definecolor{blue}{rgb}{0.0,0.0,0.75}
\newcolumntype{.}{D{.}{.}{4}}
\newcolumntype{,}{D{.}{.}{2}}
\newcolumntype{;}{D{.}{.}{1}}
\newcommand{\nodata}{$\cdot\cdot\cdot$}
\newcommand{\lesssim}{{\lower-1.2pt\vbox{\hbox{\rlap{$<$}\lower5pt\vbox{\hbox{$\sim$}}}}}}
\newcommand{\gtrsim}{{\lower-1.2pt\vbox{\hbox{\rlap{$>$}\lower5pt\vbox{\hbox{$\sim$}}}}}}
\def\clap#1{\hbox to0pt{\hss#1\hss}}

\title[CO dissociation in 47 Tucanae]{ALMA reveals sunburn: CO dissociation around AGB stars in the globular cluster 47 Tucanae}
\author[I. McDonald et al.]{I.~McDonald$^{1}$\thanks{E-mail: mcdonald@jb.man.ac.uk}, A.~A.~Zijlstra$^{1}$, E.~Lagadec$^{3,4}$, G.~C.~Sloan$^{3}$, M.~L.~Boyer$^{2}$, \newauthor  M.~Matsuura$^{5,6}$, R.~J.~Smith$^{1}$, C.~L.~Smith$^{1}$, J.~A.~Yates$^{5}$, J.~Th.~van~Loon$^{7}$, \newauthor O.~C.~Jones$^{1,8}$, S.~Ramstedt$^{9}$, A.~Avison$^{1}$, K.~Justtanont$^{10}$, H.~Olofsson$^{10}$, \newauthor J.~A.~D.~L.~Blommaert$^{11}$, S.~R.~Goldman$^{7}$, M.~A.~T.~Groenewegen$^{12}$\\
$^{1}$Jodrell Bank Centre for Astrophysics, Alan Turing Building, School of Physics and Astronomy, The University of Manchester, \\Oxford Road, Manchester, M13 9PL, UK\\
$^{2}$Observational Cosmology Lab, Code 665, NASA Goddard Flight Center, Greenbelt, MD 20771, USA\\
$^{3}$The Cornell Center for Astrophysics and Planetary Science, Cornell University, Ithaca, NY 14853-6801, USA\\
$^{4}$Observatoire de la C\^ote d'Azur, Boulevard de l'Observatoire, CS 34229, F 06304 Nice Cedex 4, France\\
$^{5}$Department of Physics and Astronomy, University College London, Gower Street, London, WC1E 6BT, UK\\
$^{6}$School of Physics and Astronomy, Cardiff University, Queen's Buildings, The Parade, Cardiff, CF24 3AA, UK\\
$^{7}$Astrophysics Group, Lennard-Jones Laboratories, Keele University, ST5 5BG, UK\\
$^{8}$STScI, 3700 San Martin Drive, Baltimore, MD 21218, USA\\
$^{9}$Department of Physics and Astronomy, Uppsala University, Box 516, SE-75120 Uppsala, Sweden\\
$^{10}$Chalmers University of Technology, Dept.\ of Earth \& Space Science, Onsala Space Observatory, S-439 92 Onsala, Sweden\\
$^{11}$Astronomy and Astrophysics Research Group, Department of Physics and Astrophysics, Vrije Universiteit Brussel, Pleinlaan 2,\\ \quad 1050 Brussels, Belgium\\
$^{12}$Koninklijke Sterrenwacht van Belgi\"e, Ringlaan 3, B--1180 Brussels, Belgium\\
}

\begin{document}

\date{Accepted 9999 December 32. Received 9999 December 32; in original form 9999 December 32}

\pagerange{\pageref{firstpage}--\pageref{lastpage}} \pubyear{9999}

\maketitle

\label{firstpage}

\begin{abstract}

ALMA observations show a non-detection of carbon monoxide around the four most luminous asymptotic giant branch (AGB) stars in the globular cluster 47 Tucanae. Stellar evolution models and star counts show that the mass-loss rates from these stars should be $\sim$1.2--3.5 $\times$ 10$^{-7}$ M$_\odot$ yr$^{-1}$. We would na\"ively expect such stars to be detectable at this distance (4.5 kpc). By modelling the ultraviolet radiation field from post-AGB stars and white dwarfs in 47 Tuc, we conclude CO should be dissociated abnormally close to the stars. We estimate that the CO envelopes will be truncated at a few hundred stellar radii from their host stars and that the line intensities are about two orders of magnitude below our current detection limits.  The truncation of CO envelopes should be important for AGB stars in dense clusters. Observing the CO (3--2) and higher transitions and targeting stars far from the centres of clusters should result in the detections needed to measure the outflow velocities from these stars.
\end{abstract}




\section{Introduction}
\label{IntroSect}
\label{IntroDrivingSect}

\begin{table*}
 \centering
 \begin{minipage}{160mm}
  \caption{Published properties of the observed stars.}
\label{StarsTable}
  \begin{tabular}{lcccccccccccccc}
  \hline\hline
   \multicolumn{1}{c}{ID} & \multicolumn{1}{c}{RA}	& \multicolumn{1}{c}{Dec} 	& \multicolumn{1}{c}{$r$}	&
	\multicolumn{1}{c}{$<\!v_{\rm LSR}\!>$}	&\multicolumn{1}{c}{$\delta v$}	& \multicolumn{1}{c}{$\delta V$}	& \multicolumn{1}{c}{$P$}	&
	\multicolumn{2}{c}{$L$}	&\multicolumn{2}{c}{$T_{\rm eff}$} & \multicolumn{1}{c}{$\phi$} \\
   \multicolumn{1}{c}{\ } & \multicolumn{1}{c}{(J2000)} & \multicolumn{1}{c}{(J2000)}	& \multicolumn{1}{c}{($^\prime$)}&
	\multicolumn{1}{c}{(km s$^{-1}$)}& \multicolumn{1}{c}{(km s$^{-1}$)}	& \multicolumn{1}{c}{(mag)}		& \multicolumn{1}{c}{(days)}	&
	\multicolumn{2}{c}{(L$_\odot$)}	&\multicolumn{2}{c}{(K)} & \multicolumn{1}{c}{\ }\\
 \hline
Notes	& (1)         & (1)          & (2)  & (3)    & (3)  & (3)  & (3) & (4)  & (5)  & (4)  & (5)        & (5) \\
V1	& 00 24 12.65 & --72 06 39.9 & 1.87 & --14.7 & 10   & 4.03 & 221 & 4824 & 4760    & 3623 & 3410    & 0.90 \\
V2	& 00 24 18.56 & --72 07 59.0 & 3.26 & --12.7 & 12   & 2.78 & 203 & 3031 & 4470    & 3738 & 3620    & 0.35 \\
V3	& 00 25 15.96 & --72 03 54.8 & 5.49 & --35.2 & 11.5 & 4.15 & 192 & 2975 & 4590    & 3153 & 3540    & 0.18 \\
V8	& 00 24 08.59 & --72 03 54.9 & 0.99 & --32.7 & 8    & 1.6  & 155 & 3583 & \nodata & 3578 & \nodata & \nodata \\
\hline
\multicolumn{13}{p{0.97\textwidth}}{References: (1) \citet[][2MASS]{CSvD+03}; (2) 2MASS offset from cluster core (00 24 05.67, --72 04 52.6); (3) average and semi-amplitude of radial velocities from \citet{LW05,LWH+05} with a --8.7 km s$^{-1}$ correction from heliocentric velocity, plus $V$-band magnitude pulsation  amplitudes (full range) and periods; (4) \citet{MBvLZ11}, based on multi-wavelength literature photometry; (5) \citet{LNH+14}, based on instantaneous measurements at the given pulsation phase.}\\
\hline
\end{tabular}
\end{minipage}
\end{table*}

\begin{table*}
 \centering
 \begin{minipage}{160mm}
  \caption{Predicted properties of the observed stellar winds.}
\label{WindsTable}
  \begin{tabular}{lccccccccccccc}
  \hline\hline
   \multicolumn{1}{c}{ID} & $\dot{M}_{\rm IR}$& $v_{\rm exp}^{\rm dust}$	& $T_{\rm dust}^{\rm max}$ & Dust & $\dot{M}_{\rm Reimers}$ & $\dot{M}_{\rm SC05}$\\
   \multicolumn{1}{c}{\ } & (M$_\odot$ yr$^{-1}$) & km s$^{-1}$	& (K)                   & composition & (M$_\odot$ yr$^{-1}$) & (M$_\odot$ yr$^{-1}$) &\\
 \hline
Notes	& (1)	                  & (1,3) & (1)  & (1)                            & (2)						& (2)				& \\
V1	& 2.1 $\times$ 10$^{-6}$ & 4.0	&  900 & Silicate, metallic iron        & 2.9$^{+1.5}_{-1.6}$ $\times$ 10$^{-7}$	& 1.1$^{+0.5}_{-0.6}$ $\times$ 10$^{-6}$& \\
V2	& 1.2 $\times$ 10$^{-6}$ & 3.8	&  900 & Silicate, metallic iron        & 2.1$^{+1.2}_{-1.1}$ $\times$ 10$^{-7}$	& 6$^{+5}_{-4}$ $\times$ 10$^{-7}$	& \\
V3	& 0.9 $\times$ 10$^{-6}$ & 3.2	& 1000 & Metallic iron only             & 2.4$^{+1.2}_{-1.0}$ $\times$ 10$^{-7}$	& 7$^{+5}_{-4}$ $\times$ 10$^{-7}$	& \\
V8	& 1.5 $\times$ 10$^{-6}$ & 4.0	&  900 & Silicate, alumina, metallic iron, oxides & 1.9$^{+0.5}_{-0.5}$ $\times$ 10$^{-7}$	& 1.0$^{+0.2}_{-0.3}$ $\times$ 10$^{-6}$& \\
\hline
\multicolumn{13}{p{0.97\textwidth}}{References: (1) mass-loss rates, wind expansion velocities, and dust condensation temperatures and mineralogies, based on modelling dust emission in infrared spectra, from \citet{MBvLZ11}; (2) based on \citet{MZ15b}: minimum expected mass-loss rates following \citet{Reimers75} with $\eta = 0.477 \pm 0.070 ^{+0.050}_{-0.062}$ and \citet[][hereafter SC05]{SC05} with $\eta = 0.172 \pm 0.024 ^{+0.018}_{-0.023}$, both assuming $M = 0.55$ M$_\odot$; (3) expansion velocities neglect any momentum input other than radiation pressure on dust, based on \citet{NIE99}.}\\
\hline
\end{tabular}
\end{minipage}
\end{table*}

Our understanding of how baryonic matter is recycled from metal-poor stars back into their host environment currently faces two major problems: how is mass actually lost from the stars and what happens to it when it is returned to the interstellar medium (ISM)?

Mass loss from giant stars follows a two-stage process. Chromospheric or magneto-acoustically-driven mass loss occurs during a star's red giant branch (RGB) and early asymptotic giant branch (AGB) evolution, when warm ($\sim$6000 K) plasma is ejected from the surface (e.g.\ \citealt{DHA84,LCW+98,LD00,SC05}). Although initially faster, this wind slows to $\sim$10 km s$^{-1}$ at the luminosity of the RGB tip \citep{MvL07,Groenewegen14}. Later, $\kappa$-mechanism pulsations can levitate material from the star, enhancing the mass-loss rate \citep{Wood79,Bowen88b}. Strongly linked to this pulsation is the formation of dust in the denser stellar atmosphere \citep[e.g.][]{BH12,MZB12,MZS+14}. Radiation pressure on this dust drives the stellar wind in a cooler outflow, also of typically $\sim$10 km s$^{-1}$, rising with stellar luminosity to $\sim$20 km s$^{-1}$ (e.g.\ \citealt{LFOP93,MvLM+04}). Although dust production begins at luminosities below the RGB tip \citep{SMM+10,MBvL+11,MJZ11,MZB12}, sufficient radiation pressure to drive the wind is only achieved once the star reaches $\sim$1000--5000 L$_\odot$ (e.g.\ \citealt{WlBJ+00,RMF+10,IM11,MZS+14}).

However, the expansion velocities of metal-poor stellar winds are poorly known. The stars are smaller compared to metal-rich stars at the same luminosity, so pulsations are thought to be weaker \citep{KB95}. The stellar surface resides deeper in the gravitational potential, so material must be levitated further before reaching temperatures where dust can condense. We also expect the dust:gas ratio will decrease, meaning that the associated mass-loss rate and velocity enhancements will also decrease. The transition between a chromospheric and pulsation-enhanced, dust-driven wind should consequently happen at a higher luminosity \citep{WlBJ+03,MBvLZ11,MZB12}. Without measurement of these expansion velocities, however, these theories remain untested. The wind expansion velocity is an important parameter in determining mass-loss rates from radiative transfer modelling of infrared excesses, which is a valuable method of determining the mass of both dust and gas ejected by giant stars, and the only method currently feasible outside our Galaxy \citep[e.g.][]{MBvLZ11,BSvL+11,BSR+12,SML+12,JvLKM13,JMR+15,BMB+15}.

Once material leaves the star it is reprocessed, first by the interstellar radiation field (ISRF) and then by interstellar shocks. Other unknowns then become important in determining how quickly dust is destroyed, molecules are dissociated, and atoms ionised. Contributing factors include the star-formation history of the population (setting the UV flux from high-mass main-sequence, low-mass horizontal-branch and post-AGB stars), the amount of interstellar dust shielding, and the local stellar density. There is a growing realisation that the enrivonment in which stars lie can have profound effects on their molecular and mineralogical yields \citep{ZPH15}.

Nowhere is this more strongly felt than globular clusters. The preparation of this work sparked the theory that ionisation should exert a powerful control over the interstellar environment of globular clusters \citep{MZ15a}. Very strong ISRFs should be generated by post-AGB stars and the hottest white dwarfs, but the short-lived nature of these objects means the ISRF is constantly varying. Not only should these sources be capable of dissociating and ionising the intra-cluster medium (ICM), but they also provide the ICM with enough thermal energy that it overflows the globular cluster, preventing further star formation.

We now turn our attention to the circumstellar environments of globular cluster stars. Many of the above issues can be addressed by observing \mbox{(sub-)mm} CO lines emanating from the winds of stars in globular clusters. CO is one of the first molecules to form as the extended atmosphere of the star is ejected, and it survives until its dissociation by the ISRF. This dissociation is mainly by photo-absorption in molecular lines, hence even in the absence of dust, self-shielding can be important in protecting CO in stellar winds. Observationally, the widths of \mbox{(sub-)mm} CO lines provide a direct measurement of the wind-expansion velocity, while their strength provides an independent estimate of mass-loss rate. Their strengths are also affected by the characteristic radius from the star at which CO is dissociated by the ISRF \citep{MGH88}. Hence, by comparing the strength of CO lines from stars with known mass-loss rates to model predictions, we can quantify the strength of the ISRF to which they are subjected.

In this paper, we report on observations with the Atacama Large Millimetre Array (ALMA) of the four brightest stars in 47 Tucanae. Section \ref{TopObsSect} describes these observations, and defines upper limits to the non-detections we find. Section \ref{CODissSect} models the expected mass-loss rate from the star, the size of the CO envelope, and calculate the expected strength of the CO transitions. This is done under the assumption that the ISRF is strong dissociating the CO envelope. Results are discussed in Section \ref{DiscSect}. Alternative scenarios which might remove the CO envelopes of our target stars are explored in the Appendix.


\section{Observations}
\label{TopObsSect}

\subsection{Target selection}
\label{Intro47TucSect}

Among the closest and most populous globular clusters is the intermediate-metallicity cluster 47 Tuc (NGC 104; 4.5 kpc, [Fe/H] = --0.72 dex, $M$ = 1.5 $\times$ 10$^6$ M$_\odot$; \citealt{GZP+02,Harris10}). This cluster was chosen partly because its stars show a relatively small spread in chemical abundances compared to other clusters of similar size \citep{WCMvL10,GLS+13,CPJ+14,CKB+14,DKB+14,LNH+14,TSA+14,JMP+14}, and partly because it contains a well-studied set of luminous, dusty AGB stars.

The cluster's variable stars have a unique, long and well-covered history \citep{Pickering1894}. Infrared photometry and spectra have been published at many wavelengths over many epochs \citep{GF73,OSA+97,RJ01,OFFPR02}. Mid-infrared spectroscopy of the brightest variables \citep{LPH+06,vLMO+06,MBvLZ11} show a mixture of silicate and alumina dust features, and a potential featureless contribution from metallic iron dust \citep[cf.][]{MSZ+10}. This combination has led to several estimates of their dust-production rates \citep{ORF+07,BvLM+10,ORF+10,MBvL+11,MBvLZ11,MSS+12}. High-resolution optical spectroscopy of moderately bright giants suggest relatively slow outflows from the stellar chromospheres of $\sim$10 km s$^{-1}$ \citep{MvL07}. Near-infrared spectroscopy \citep{LNH+14} shows that the brightest stars have abundances similar to other stars in the cluster. They have not experienced substantial third dredge-up, and perhaps have not experienced any third dredge-up episodes.

We present sub-mm CO observations of the four most luminous, most evolved stars in 47 Tuc. They are denoted by \citet{SawyerHogg73} as V1, V2, V3 and V8. Previously published parameters of these stars from the above references are given in Tables \ref{StarsTable} and \ref{WindsTable}. In particular, Table \ref{WindsTable} gives the mass-loss rate ($\dot{M}_{\rm IR}$) and wind velocity ($v_{\rm exp}^{\rm dust}$) predicted for a purely radiation-driven wind. \citet{MBvLZ11} derived these on the basis of radiative transfer modelling of their infrared spectra using the {\sc dusty} code \citep{NIE99} with a radiation-driven wind ({\tt density type = 3}). In all four cases, the spectral energy distributions were best modelled with a mostly metallic iron wind, with some stars having contributions from amorphous and crystalline silicates, aluminium oxide and iron oxide. Table \ref{WindsTable} also gives the mass-loss rate expected for a purely magneto-acoustically driven wind, as derived from semi-empirical scaling laws applied to the cluster's other stars ($\dot{M}_{\rm Reimers}$, $\dot{M}_{\rm SC05}$; see Section \ref{MdotSect}).

\begin{table}
 \centering
  \caption{Properties of the observed stellar winds derived from the upper limit to the CO(2$\rightarrow$1) line flux.}
\label{Winds2Table}
  \begin{tabular}{lccc}
  \hline\hline
   \multicolumn{1}{l}{ID} & CO(2$\rightarrow$1) & R.M.S.\ Noise & $\sigma$ \\
   \multicolumn{1}{l}{\ } & (mJy\,km\,s$^{-1}$)   & (mJy\,km\,s$^{-1}$) \\
 \hline \vspace{1pt}
V1	& $<$183 &  85 & 2.1 \\
V2	& $<$168 & 114 & 1.4 \\
V3	& $<$170 &  99 & 1.7 \\
V8	& $<$162 &  70 & 2.3 \\
\hline
\multicolumn{4}{p{0.45\textwidth}}{Notes: Based on CO (2$\rightarrow$1) line flux. The root-mean squared (r.m.s.) noise is the quadrature-summed average flux of a region of identical width, offset by +100 km s$^{-1}$ from the `detection'. The significance ($\sigma$) is the multiple of the r.m.s.\ noise by which the line `detection' is above the mean.}\\
\hline
\end{tabular}
\end{table}

\subsection{ALMA \& APEX observations}
\label{ObsALMASect}

\begin{figure}
\centerline{\includegraphics[height=0.47\textwidth,angle=-90]{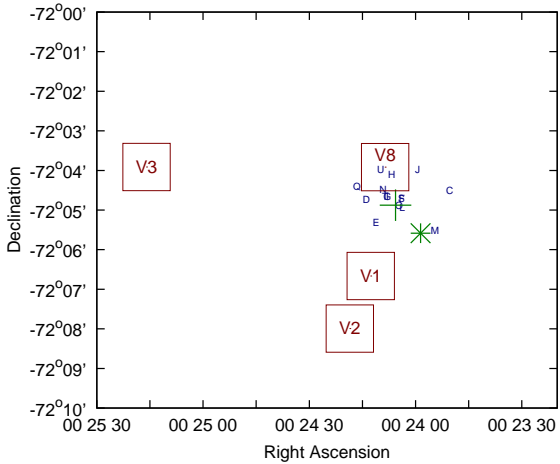}}
\caption{Map of imaged regions (squares), with the cluster centre (green $+$ symbol) and pulsars (small blue letters, from \citealt{FCK+03}) shown. The large, green asterisk marks the only significant UV source: an optically bright post-AGB star.}
\label{MapFig}
\end{figure}

Observations of 47 Tuc V1, V2, V3 and V8 (see Figure \ref{MapFig}) were carried out using ALMA in receiver band 6 on the night of 2013 Nov 05 over a 1.05 h integration. Observations were centred on the $^{12}$C$^{16}$O $J = 2 \rightarrow 1$ transition at 230.538 GHz, and were carried out at high precipitable water vapour (2.8--3.5 mm) but otherwise good conditions. The correlator output consisted of 3840 channels of 244.141 kHz each. The data were processed following the standard ALMA quality assurance calibration and imaging processes. Neptune was used for amplitude calibration and the quasars J1924--291 and J0102--7546 for bandpass and phase calibration, respectively.

During the ALMA observations the majority of antennae (26 of 29) were positioned within $\sim$250m of the array centre; the remaining three telescopes were at larger distances ($\sim$530m, $\sim$790m and $\sim$1014m). At the imaging stage of data reduction the visibilities from these three antenna were downweighted\footnote{This provides a trade-off between competing noise sources: noise is increased by the larger solid angle within the beam, but additional noise from the poorer phase calibration on longer baselines is reduced.}. This was accomplished using the {\tt uvtaper} parameter set to {\tt True} and the {\tt outaper} option set to 1.5$^{\prime\prime}$ in CASA's {\sc clean} task \citep{CASA}. In addition to this a restoring beam of $2^{\prime\prime} \times 2^{\prime\prime}$ was also applied at the CLEANing stage. These techniques were used to generate images with a beam more representative of the antennae in the centre of the array. Imaging additional to that provided with the ALMA data release was conducted in CASA using the {\sc clean} routine to create two data cubes for each source, one with a binning of ten spectral channels (channel width: 3.2 km s$^{-1}$ bin$^{-1}$) and one with two-channel binning (0.64 km s$^{-1}$ bin$^{-1}$). The pixel scale in both cubes is 0.15$^{\prime\prime}$ pixel$^{-1}$.

No line or continuum source is visible in any of the maps to the depth of the observations. The standard deviation of the noise in the final images is $\sim$14 mJy beam$^{-1}$ per 0.64 km s$^{-1}$ channel pair for V1, V3 and V8, and $\sim$16 mJy beam$^{-1}$ per channel pair for V2. The shortest baseline (17 m) corresponds to a maximum detectable object size of 15$\farcs$7, though the CO shells are expected to be unresolved ($\approx$1$^{\prime\prime}$; Section \ref{COExpectSect}). On-source spectra were extracted using a 3$\times$3 pixel box surrounding the central source.

Figure \ref{SpecFig} shows the two-channel-averaged spectra around the velocity of the cluster. The maximum two-channel flux for each source is 30, 33, 36 and 37 mJy beam$^{-1}$ for V1, V2, V3 and V8, respectively. Each spectrum was examined and found to have noise following a Gaussian distribution, both on each source and at various points away from each source. On-source standard deviations in the spectra are 10.2, 11.9, 11.6 and 11.2 mJy beam$^{-1}$ per 0.64 km s$^{-1}$ bin for V1, V2, V3 and V8, respectively.

Additional data were obtained from the Atacama Pathfinder Experiment (APEX) telescope. A four-hour integration on 47 Tuc V1 was made in the 345 GHz $^{12}$C$^{16}$O $J = 3 \rightarrow 2$ transition (project identifier: O-092.F-9327A).  Standard reduction of the APEX data was performed at Onsala Space Observatory before receipt of the data. No line was detected, with a limiting peak flux of $\sim$0.1 Jy over a typical 10 km s$^{-1}$ bin from the 17$\farcs$3 beam.

\subsection{Results \& comparison to previous observations}
\label{ObsResultsSect}

\begin{figure}
\centerline{\includegraphics[height=0.47\textwidth,angle=-90]{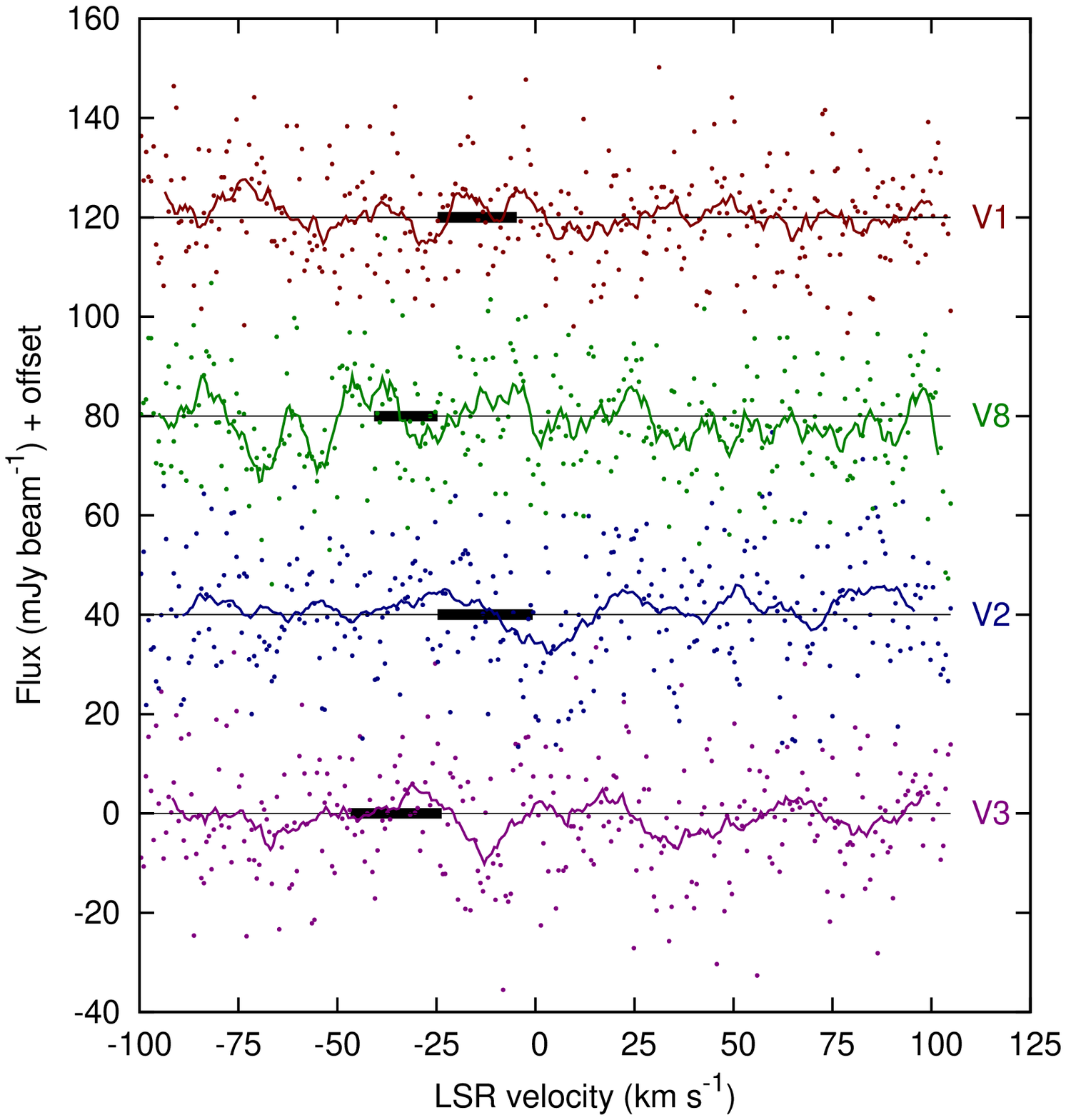}}
\caption{On-source ALMA spectra of the four observed sources (points), offset by an arbitrary flux level with zero flux lines shown. The cluster mean velocity is --26.7 km s$^{-1}$ \citep{Harris10} and the range of radial velocities over the pulsation cycle are shown as thick horizontal lines \citep{LWH+05}. Lines show the spectra, boxcar-smoothed by 11, 20, 15 and 7.4 km s$^{-1}$ for V1, V2, V3 and V8, respectively. This smoothing corresponds to the upper mass-loss rate limits quoted in Table \ref{Winds2Table}.}
\label{SpecFig}
\end{figure}

Neither the ALMA nor the APEX data showed any convincing indication of spectrally resolved CO emission (e.g.\ departure from Gaussian noise centred around zero flux).

The only previous significant CO observations of the cluster are by \citet{OGFFP97}, who claim a weak CO (1$\rightarrow$0) detection around the cluster's velocity with a signal-to-noise ratio of 3. We interpret their peak fluxes as 1.6 Jy, assuming an aperture efficiency for the Swedish--ESO Sub-millimetre telescope\footnote{http://www.apex-telescope.org/sest/html/telescope-instruments/telescope/index.html.} (SEST) of 27 Jy K$^{-1}$. Origlia et al.'s observing position is not quoted, but assumed to be $\alpha \approx 00$ 24 22, $\delta \approx -72$ 05 48, following their offset from \citet{KG95}. This is 68$^{\prime\prime}$ from V1, compared to a full-width half-maximum SEST beam width of 45$^{\prime\prime}$. In neither dataset do we recover any source comparable to the CO (1--0) detection of \citet{OGFFP97}, though the beam footprint of the SEST observations extends considerably beyond both the ALMA primary beam of 15$\farcs$7 and the APEX primary beam of 17$\farcs$3. 

The CO emission is expected to be spatially unresolved by ALMA and the mass-loss rate is expected to be sufficiently low that the lines will be optically thin. An unresolved, optically thin line will have a rectangular (`boxcar') spectral profile. To identify whether a low-contrast, spectrally resolved feature is present in our data, we smoothed our data using a boxcar function (running average). The boxcar was given different central velocities, from $v_{\rm LSR} = -50$ to 0 km s$^{-1}$, and different widths, from 1.27 km s$^{-1}$ to 85 km s$^{-1}$ (4--133 channels).

Typical expansion velocities for dust-producing stars in the Milky Way are $\sim$10--20 km s$^{-1}$, and are typically $\sim$10 km s$^{-1}$ for stars producing little dust \citep[e.g.][]{SO01,DTJ+15}. We presume the expansion velocity is $v_{\rm exp} < 20$ km s$^{-1}$, and that the centre of the emission line is Doppler shifted by no more than 5 km s$^{-1}$ from the average radial velocity of the star. The limits to the velocity-integrated CO line strength are typically below $<$200 mJy km s$^{-1}$, with a typical 1$\sigma$ uncertainty being $\sim$100 mJy km s$^{-1}$ (see Table \ref{Winds2Table}). The boxcar-smoothed spectra producing these maximum fluxes are shown in Figure \ref{SpecFig}.

The mean photospheric radial velocities and their pulsation-induced variation are listed in Table \ref{StarsTable}\footnote{Note that the velocities published by \citet{LWH+05} are given as heliocentric velocities (T.\ Lebzelter, private communication), although this is not explictly stated in the text. A correction of --8.7 km s$^{-1}$ has been applied.}. The highest flux in the V1 and V8 spectra is within 10 km s$^{-1}$ of the mean photospheric velocities of the stars as published by \citet{LWH+05}. These may indicate a tentative detection of the CO line, but we do not consider them reliable as they are only at the 1.3 and 2.1$\sigma$ confidence levels, respectively.


\section{The expected CO line flux}
\label{CODissSect}

\subsection{The CO flux under Galactic ISRF condtions}

Our observations were carried out under the na\"ive assumption that the envelopes of our target stars behave as Galactic stars do. In this section, we show that we could expect a similar Galactic star to be observable at the distance of 47 Tucanae. This provides a baseline to which we can compare our final model.

\citet{Olofsson08} provides an estimate of the peak CO(2--1) line flux from an evolved star, namely:
\begin{equation}
S_{\rm CO(2-1)} \approx 6 \left(\frac{\dot{M}}{10^{-6}}\right)^{1.2} \left(\frac{15}{v_{\rm e}}\right)^{1.6} \left(\frac{f({\rm CO})}{10^{-3}}\right)^{0.7} \left(\frac{1}{D}\right)^{2} {\rm Jy}
\end{equation}
where $\dot{M}$ is the mass-loss rate in M$_\odot$ yr$^{-1}$, $v_{\rm e}$ is the expansion velocity in km s$^{-1}$, $f_{\rm CO}$ is the ratio of CO to H$_2$, and $D$ is the distance in kpc ($D = 4.5$ kpc). From Table \ref{WindsTable}, we can modestly expect $\dot{M} \approx 2 \times 10^{-7} - 2 \times 10^{-6}$ M$_\odot$ yr$^{-1}$, depending on the star and the method used. If we assume $\dot{M} > 3 \times 10^{-7}$ M$_\odot$ yr$^{-1}$ (appropriate for V1), a canonical $v_{\rm e} = 10$ km s$^{-1}$ and $f_{\rm CO} = 4 \times 10^{-5}$ (based on scaling a canonical 2 $\times$ 10$^{-4}$ \citep{RSOL08} by the metallicity of 47 Tuc, which is a fifth of solar metallicity), we obtain a peak flux of $>$14 mJy. Assuming a boxcar profile gives an integrated intensity of $>$280 mJy km s$^{-1}$, which is still greater than our upper limit.

\citet{Olofsson08} also predicts that the CO(3--2) flux should be around twice as strong. A less conservative value of $\dot{M} = 10^{-6}$ M$_\odot$ yr$^{-1}$ produces 1200 mJy km s$^{-1}$ for the CO(2--1) line, implying 2400 mJy km s$^{-1}$ for the CO(3--2) line, or 1.2$\sigma$ in our APEX observation.

Much depends on the exact values used in the equation above, as well as the accuracy of the formula itself. Stars in globular clusters exist in environments unlike the well-studied stars in the Solar Neighbourhood. In particular, it has recently been shown that the ISRF is considerably stronger and harder \citep{MZ15a}. This will have a substantial impact on the size of the CO envelope and the line flux emitted from it. In this section, we re-evaluate the expected CO line fluxes from these stars by creating a photochemical dissociation model for V1. We consider other factors peculiar to globular clusters in the Appendix, but find that they have little impact.

\subsection{Expected mass-loss rates}
\label{MdotSect}

\subsubsection{Expected mass-loss rate from chromospheric physics}
\label{ReimersSect}

In general, mass-loss rates of stars without dust-driven winds are comparatively well determined and have been parameterised into scaling relations. Two common formalisms which are frequently adopted are the emprical formalism of \citet{Reimers75} and its semi-emprical modification by SC05. In these formalisms, the mass-loss rate is scaled by an efficiency factor, $\eta$. In terms of observable units, they can be written as:
\begin{equation}
\dot{M}_{\rm R} = 4 \times 10^{-13} \eta_{\rm R} \frac{L^{1.5}}{M T_\ast^2},
\label{ReimersEq}
\end{equation}
and:
\begin{equation}
\dot{M}_{\rm SC} = 4 \times 10^{-13} \eta_{\rm SC} \frac{L^{1.5}}{M T_\ast^2} \!\! \left( \!\frac{T_\ast}{0.6925}\! \right)^{\!\!\!3.5} \!\! \left(\!1\!+\!\frac{L}{4300MT_\ast^4} \! \right),
\label{SCEq}
\end{equation}
where the luminosity, mass and temperature ($L$, $M$, $T_\ast$) are all scaled to solar units and the mass-loss rate is in M$_\odot$ yr$^{-1}$. Throughout a large sample of Milky Way globular clusters, \citet{MZ15b} compared the mass lost between the main-sequence turn-off and the horizontal branch and compared it to theoretical isochrones at fixed $\eta$. This fixes $\eta$ around the point where mass loss is strongest, close to the RGB tip. From these, the following efficiency parameters were derived for the median stars in globular clusters:
\begin{eqnarray}
\eta_{\rm R} &=& 0.477 \pm 0.070 ^{+0.050}_{-0.062} \mbox{, and} \nonumber\\
\eta_{\rm SC} &=& 0.172 \pm 0.024 ^{+0.018}_{-0.023} .
\label{EtaEq}
\end{eqnarray}
where each value is given with its respective statistical and systematic error\footnote{\citet{HKR+15} model diffusion of AGB stars in 47 Tuc, arguing for a lower $\eta$ on the RGB ($\eta_{\rm R} \approx 0.1$), and a higher $\eta$ on the AGB ($\eta_{\rm R} \approx 0.7$). This would increase the predicted mass-loss rate for our targets. Private communications with Heyl et al.\ have not identified the source of the discrepancy. \citet{LW05} also argue for a slightly smaller $\eta_{\rm R} \approx 0.33$, on the basis of the stellar period--luminosity diagram.}. From these, we derive the minimum mass-loss rates listed in Table \ref{WindsTable} of $\dot{M}_{\rm Reimers} \approx 2-3 \times 10^{-7}$ M$_\odot$ yr$^{-1}$ and $\dot{M}_{\rm SC05} \approx 7-11 \times 10^{-7}$ M$_\odot$ yr$^{-1}$. These values are appropriate for `stable' stars, without extra energy from pulsations or radiation pressure on dust. While the interplay of pulsations on transfer of magneto-acoustic energy to the chromosphere is poorly determined, we would expect these values to provide a lower limit to the mass-loss rate from pulsating, dusty AGB stars.

\subsubsection{Expected maximum rate from stellar evolutionary arguments}
\label{MaxMdotSect}

The mass lost from an AGB star during the last part of its life must not exceed the envelope mass of the star at the beginning of that period. {Low-luminosity AGB stars are not easily seperable from RGB stars. AGB stars can only be uniquely identified once they exceed the luminosity of the RGB tip.} If we can determine both the remaining lifetime of an AGB star passing the luminosity of the RGB tip, and the envelope mass it has when it does so, the quotient of the two should yield the average mass-loss rate of AGB stars above the RGB-tip luminosity.

The time spent above the RGB tip is relatively easy to compute. Our four stars lie above the RGB of 47 Tuc. For the purposes of these calculations, we also include the fifth-brightest star in the cluster. V4 is a probable AGB star at the luminosity of the RGB tip (see the Hertzsprung--Russell diagram presented by \citealt{MBvLZ11}). Like its brighter counterparts, it shows long-period pulsations and an infrared excess, consistent with circumstellar dust production. Assuming there are typically 5 $\pm$ $\sqrt{5}$ stars above the cluster's RGB tip, and taking an evolutionary rate of one star passing the RGB tip every 80\,000 years \citep{MBvLZ11}, these five stars should represent the last 0.40 $\pm$ 0.18 Myr of AGB evolution in the cluster.

The envelope mass at the luminosity of the RGB tip is more difficult to calculate. \citet{GCB+10} determine zero-age horizontal branch (ZAHB) star masses in 47 Tuc. Their ground-based measurements suggest the ZAHB star mass ranges from 0.629 M$_\odot$ to 0.666 M$_\odot$, with a median of 0.648 M$_\odot$, but they also include masses derived from \emph{Hubble Space Telescope (HST)} photometry, which range from 0.650 M$_\odot$ to 0.691 M$_\odot$, with a median of 0.674 M$_\odot$. We assume these stars will end their lives as a 0.53 M$_\odot$ white dwarf, based on measured masses of white dwarfs in other clusters \citep{RFI+97,MKZ+04,KSDR+09}. These masses imply that $\sim$0.12 M$_\odot$ or $\sim$0.16 M$_\odot$ is lost during the entire horizontal branch and AGB evolution, including that below the RGB-tip luminosity. Assuming the limiting case of losing 0.16 M$_\odot$ over 221\,000 years, the \emph{average} mass-loss rate for stars above the RGB tip of 47 Tuc must be $<$7 $\times$ 10$^{-7}$ M$_\odot$ yr$^{-1}$.

A more limiting case is reached if we can estimate the envelope mass at the luminosity of the RGB tip itself. Calculating the mass lost between the zero-age horizontal branch up the AGB to the luminosity of the RGB tip is not trivial. The evolutionary track used by \citet{MZ15b} was designed to reproduce the RGB and HB of 47 Tucanae. This track was generated with the {\sc mesa} (Modules for Experiments in Stellar Astrophysics) stellar evolution code \citep{PBD+11,PCA+13}. Mass loss was included at the rate of $\eta_{\rm R} = 0.45$, reproducing a ZAHB mass of 0.666 M$_\odot$ after 12.02 Gyr. The model loses a further 0.067 M$_\odot$ between the ZAHB and the point where the star reaches the same luminosity as the RGB tip (i.e.\ the same luminosity as V4). Taking the limiting case of the \emph{HST}-derived mass of 0.674 M$_\odot$, a total mass at the luminosity of the RGB tip of 0.607 M$_\odot$ is implied, giving an envelope mass of $\sim$0.077 M$_\odot$ and limiting the average mass-loss rate to $<$3.5 $\times$ 10$^{-7}$ M$_\odot$ yr$^{-1}$. In contrast to the mass-loss rate in the previous paragraph, this mass-loss rate requires that Reimers' mass-loss formula is appropriate for early-AGB stars.



\subsubsection{Summary}
\label{MdotSummarySect}

Reimers' scaling law\footnote{This process can be repeated with the SC05 scaling law, which predicts an average of $>$8.5 $\pm$ 4.3 $\times$ 10$^{-7}$ M$_\odot$ yr$^{-1}$ for our stars. This is inconsistent with the $<$3.5 $\times$ 10$^{-7}$ M$_\odot$ yr$^{-1}$ provided by the evolutionary analysis. This suggests the SC05 law is not appropriate for this particular situation.} provides a lower limit to the mass-loss rates for pulsating AGB stars, implying an average mass-loss rate for our four stars of $>$2.3 $\pm$ 1.1 $\times$ 10$^{-7}$ M$_\odot$ yr$^{-1}$. Stellar evolution models, pinned by the mass of HB stars (Section \ref{MaxMdotSect}) suggest an upper limit of $<$3.5 $\times$ 10$^{-7}$ M$_\odot$ yr$^{-1}$. We can therefore expect the mass-loss rate from our four observed stars to lie in the range 1.2--3.5 $\times$ 10$^{-7}$ M$_\odot$ yr$^{-1}$.

This predicted mass-loss rate is estimated to produce a CO line flux within the noise limit of our observations. However, the average predicted mass-loss rate is much less than the average of 1.4 $\times$ 10$^{-6}$ M$_\odot$ yr$^{-1}$ derived from these stars' infrared spectra\footnote{Note that these assume a dust-to-gas ratio of 1:1076. The metallicity of 47 Tuc limits the abundance of refractory elements such that the true dust-to-gas ratio should not be closer to unity than 1:778 \citep{MBvLZ11}.}, although the infrared-based rates do come with a sizeable and poorly quantifiable uncertainty. These infrared-based mass-loss rates assumed slow (2--4 km s$^{-1}$) winds, accelerated only by radiation pressure on dust; faster ($\gtrsim$10 km s$^{-1}$) winds would increase the calculated mass-loss rate.

\subsection{The interstellar radiation field (ISRF)}
\label{ISRFSect}

\begin{figure}
\centerline{\includegraphics[height=0.47\textwidth,angle=-90]{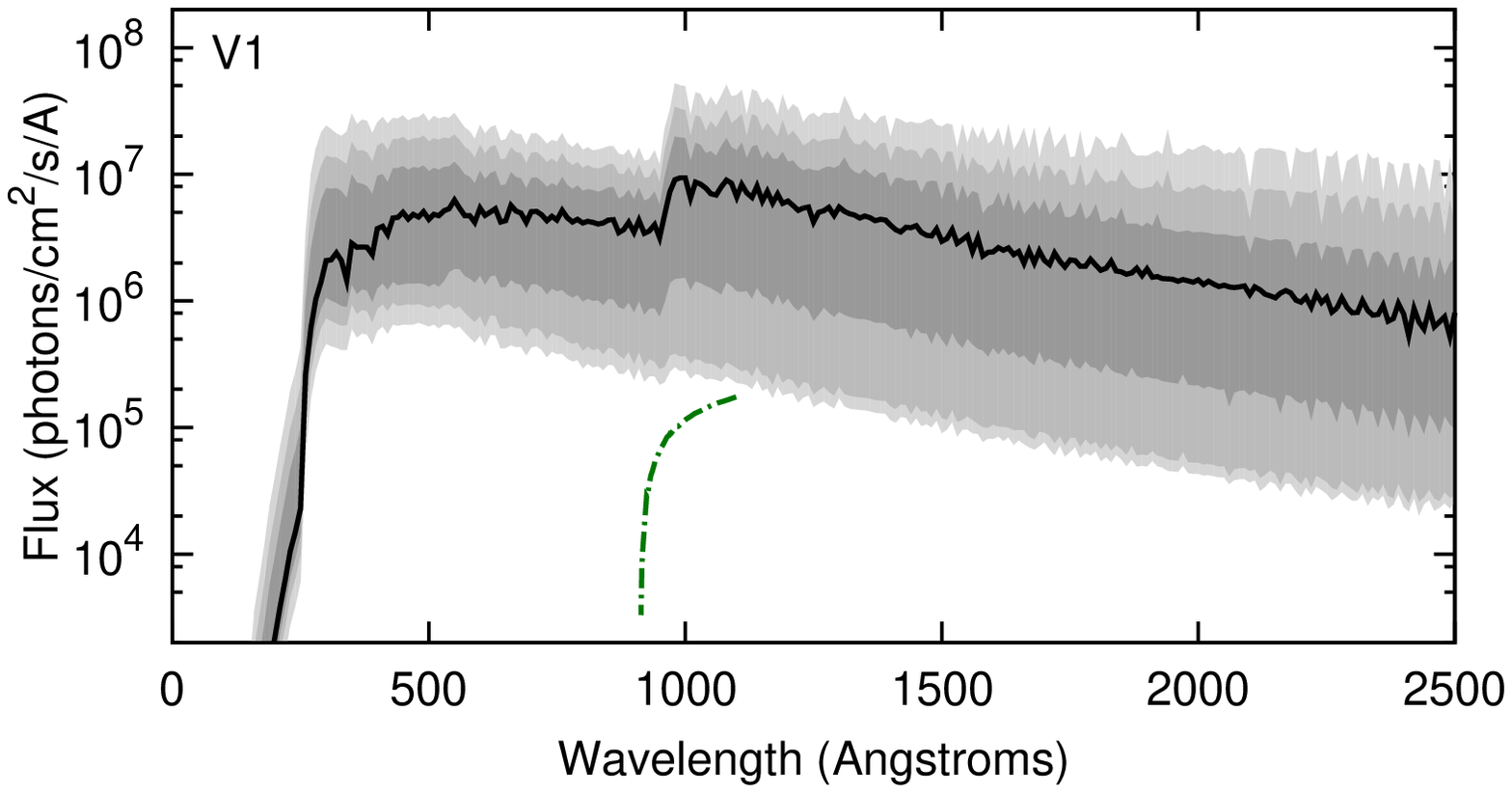}}
\centerline{\includegraphics[height=0.47\textwidth,angle=-90]{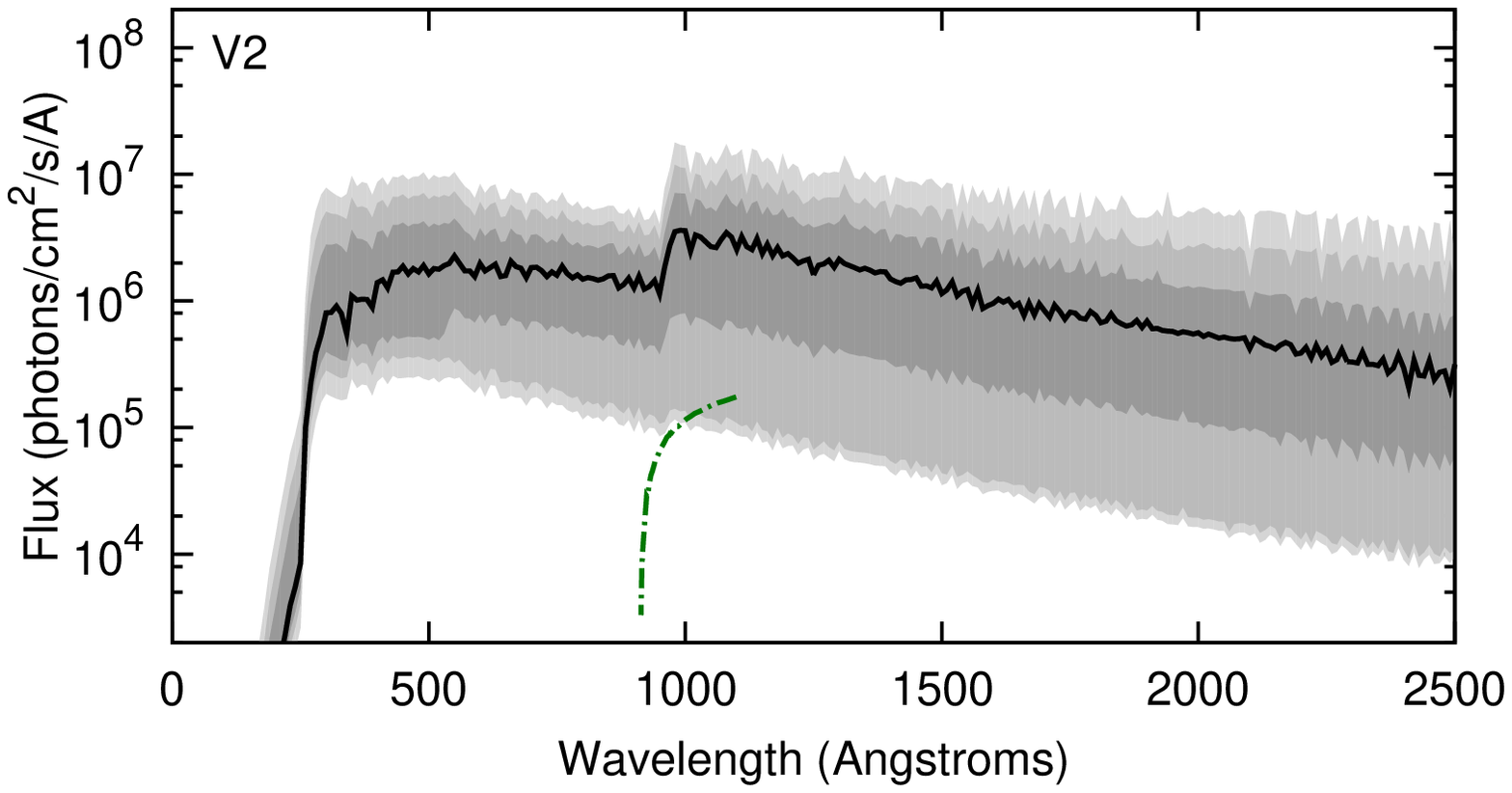}}
\centerline{\includegraphics[height=0.47\textwidth,angle=-90]{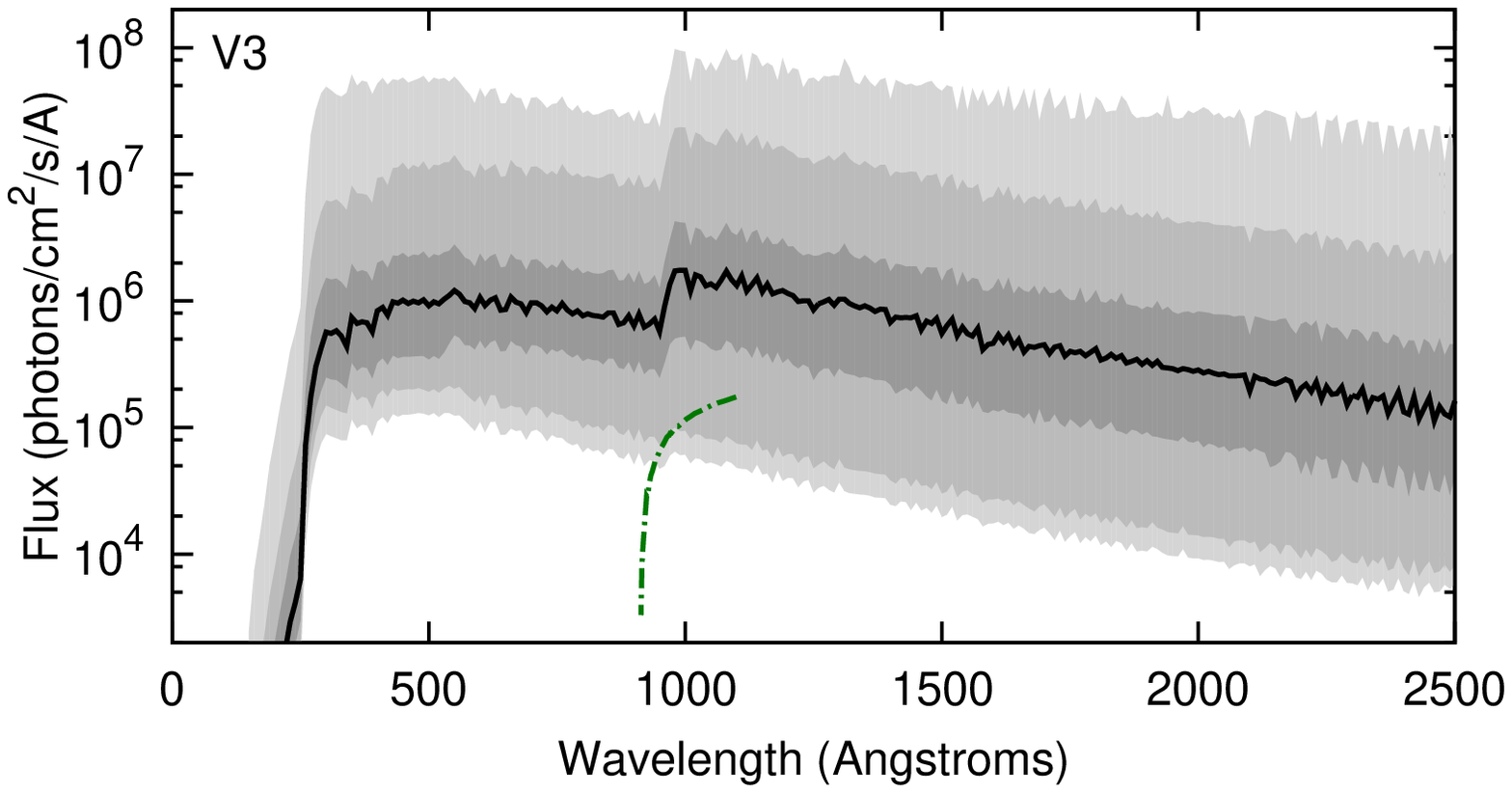}}
\centerline{\includegraphics[height=0.47\textwidth,angle=-90]{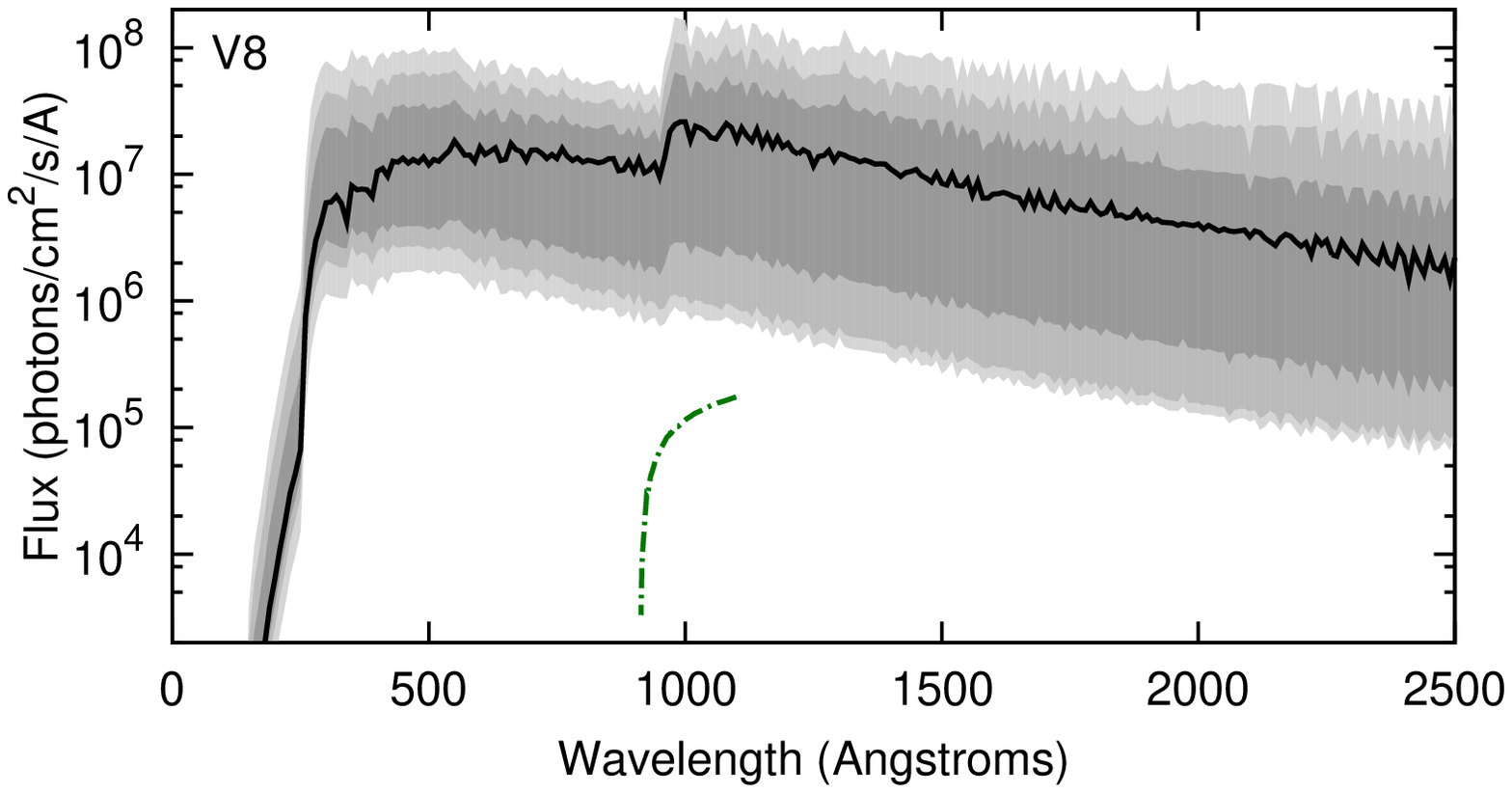}}
\caption{The ISRF estimated for each star. The line denotes the best-estimate, time-averaged value, with lightening colours of grey representing the 68.3, 95.5 and 100th centile ranges. The ISRF in the Solar Neighbourhood  (as assumed by \citealt{MGH88}) is shown as the dotted green line.}
\label{ISRFFig}
\end{figure}

\begin{table}
 \centering
  \caption{Modelled ISRF incident on each star at 1076 \AA, relative to the flux in the Solar Neighbourhood from \citet[][162\,000 photons cm$^{-2}$ s$^{-1}$ A$^{-1}$]{Jura74}. The centile probabilities approximate the best estimate, 1-$\sigma$, 2-$\sigma$ and full ranges of the modelled fluxes.}
\label{ISRFTable}
  \begin{tabular}{@{}l@{}r@{}r@{}r@{}r@{}r@{}r@{}r@{}}
  \hline\hline
   \multicolumn{1}{c}{Star} & \multicolumn{7}{c}{1076 \AA\ ISRF strength at given centiles}\\
   \multicolumn{1}{c}{\ } & \multicolumn{1}{c}{0} & \multicolumn{1}{c}{2.25} & \multicolumn{1}{c}{15.85} & \multicolumn{1}{c}{50} & \multicolumn{1}{c}{84.15} & \multicolumn{1}{c}{97.75} & \multicolumn{1}{c}{100}\\
 \hline
V1	& 1.4  & 1.8  & 7.0 & 49  & 102 & 170 & 256\\
V2	& 0.55 & 0.65 & 3.6 & 19  & 37  & 59  & 87\\
V3	& 0.31 & 0.47 & 2.6 & 9.0 & 22  & 124 & 483\\
V8	& 4.3  & 5.2  & 14  & 135 & 315 & 551 & 858\\
\hline
\end{tabular}
\end{table}


The radiation environment within globular clusters differs from that in the Solar Neighbourhood \citep{MZ15a}. The ionised hydrogen and lack of interstellar dust within globular clusters reduces the opacity for ionising photons. Ionising sources are mainly post-AGB stars and white dwarfs, rather than the O and B stars of the Solar Neighbourhood. This leads to a much harsher ISRF than is typically assumed.

The ISRF of 47 Tuc was modelled in detail by \citet{MZ15a}. The ISRF can vary considerably, as the presence of very hot post-AGB stars in the cluster is stochastic. Many of the hottest post-AGB stars and white dwarfs in the cluster remain unobserved. We must therefore rely on stellar evolution models to describe the typical ISRF within 47 Tuc and its temporal range.

To compute the ISRF impinging on each AGB star, we have created a three-dimensional model of the cluster. We begin with the catalogue from \citet{MBvL+11}, which contains the two-dimensional (RA--Dec) position of almost every post-main-sequence source in the cluster up to the early post-AGB phase. We assume this is also representative of the distribution of more-evolved, hotter post-AGB stars. For each catalogued source, we assign a physical depth within the cluster. Assuming an Earth-to-cluster distance of 4500 pc, we randomly select a source nearby in right ascension to each star and use the declination of the random source to assign a radial distance from the cluster centre for each target source. This creates a stellar distribution with the same radial density profile as the density profile in declination, approximating a spherically symmetric distribution.

We then take the post-AGB stellar evolution model of \citet{MZ15a}, following the emitted spectrum as it evolves through its upper-AGB, post-AGB and hot-white-dwarf evolution. We follow the evolution of each of our model sources in time, randomly forcing one of our sources to undergo this post-AGB evolution with an average of once per 80\,000 years (the stellar evolution rate according to \citealt{MBvL+11}). Summing the modelled flux from these sources at the locations of our four target stars, we obtain the integrated interstellar radiation field impinging upon them.

This model was run 100 times, in order to simulate different radial distributions of the radiation sources and observed targets. For each run, the evolution was followed for 10$^6$ years to quantify the time variability of the radiation.

Figure \ref{ISRFFig} shows the ISRF, averaged for each source over all runs and all times. For comparison, the figure also shows the Solar Neighbourhood ISRF used by \citet[][originally from \protect\citealt{Jura74}]{MGH88}. The radiation field in the 914--1120 \AA\ region (normally associated with CO dissociation) is typically many times higher than the \citet{MGH88} value, and it is also considerably harsher.

\subsection{Modelling the outer CO shell radius}
\label{CODissRadiusSect}

\subsubsection{A crude estimate}
\label{DissREstimateSect}

CO in the circumstellar environment has some self-shielding, but most shielding comes from circumstellar dust and H$_2$, as shown by \citet[][their figure 2]{MGH88}. At some depth the atmosphere becomes optically thick to incoming or outgoing radiation, forming the photosphere, or $\tau = 1$ layer. In mass-losing stars, particularly at short wavelengths, the $\tau = 1$ layer is wavelength dependent, depending on the shielding within the wind. Dissociation can therefore take place in an unshielded environment, if the time to intercept a UV photon is large compared to the time taken to reach the $\tau = 1$ layer, or a shielded environment if it is smaller.

If shielding is not important, the CO dissociation radius will be limited simply by the average time it takes a CO molecule to encounter an interstellar UV photon, and the CO dissociation radius around metal-poor stars should be very similar to metal-rich stars. If the shielding is important, CO will be dissociated close to the $\tau = 1$ layer, and the dissociation radius around metal-poor stars should be smaller. However, it cannot become much smaller before H$_2$ becomes the dominant shielding mechanism. In either case, we expect metallicity to only have a second-order effect on the CO dissociation radius, and that the primary effect will be the strength of the ISRF. For a Solar-Neighbourhood ISRF, we would expect the 47 Tuc stars have CO envelopes filled to $\sim$0.013--0.023 pc (Section \ref{COExpectSect}). A stronger ISRF will dissociate CO proportionally more quickly, so the dissociation radius should be inversely proportional to the ISRF strength.

The primary contribution to CO dissociation is normally the 1076 \AA\ band \citep{MGH88}, though the harder radiation field may change this in our case. The potential range in ISRF strength is large (Table \ref{ISRFTable}) and is considerably greater for the stars closer to the cluster centre (V1, V8) than stars on the cluster's outskirts (V2, V3). Taking the 1$\sigma$ ranges for the potential variation in ISRF, we compute a typical range of CO envelope outer radii of $\sim$4 $\times$ 10$^{-5}$ -- 0.008 pc, or 8--1800 AU. This does not for changes in the \mbox{(self-)shielding} of CO or the harder ISRF, and implies CO may be dissociated very close to the stellar surface ($\lesssim$10 R$_\ast$) at certain times. If the wind remains optically thin below these heights, there may be substantial CO dissociation within the dust-formation shell (2--10 R$_\ast$) around these stars. More typically, however, we would expect 90 per cent of the CO to be dissociated by $\sim$100 AU ($\sim$120 R$_\ast$) in V1, and $\sim$30 AU ($\sim$40 R$_\ast$) for V8. 

\subsubsection{A simple model}
\label{ArepoSect}

\begin{figure}
\centerline{\includegraphics[height=0.47\textwidth,angle=-90]{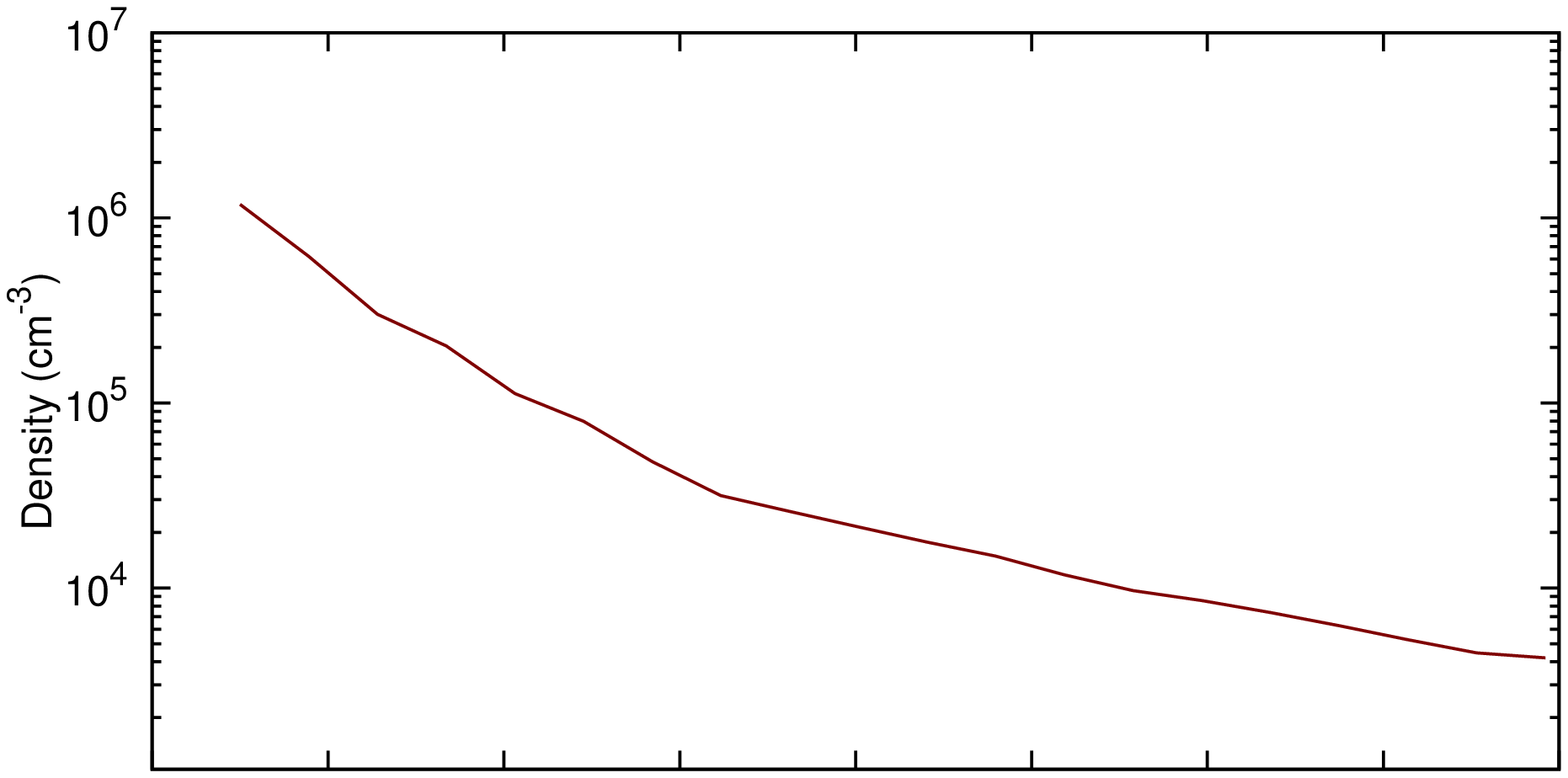}}
\centerline{\includegraphics[height=0.47\textwidth,angle=-90]{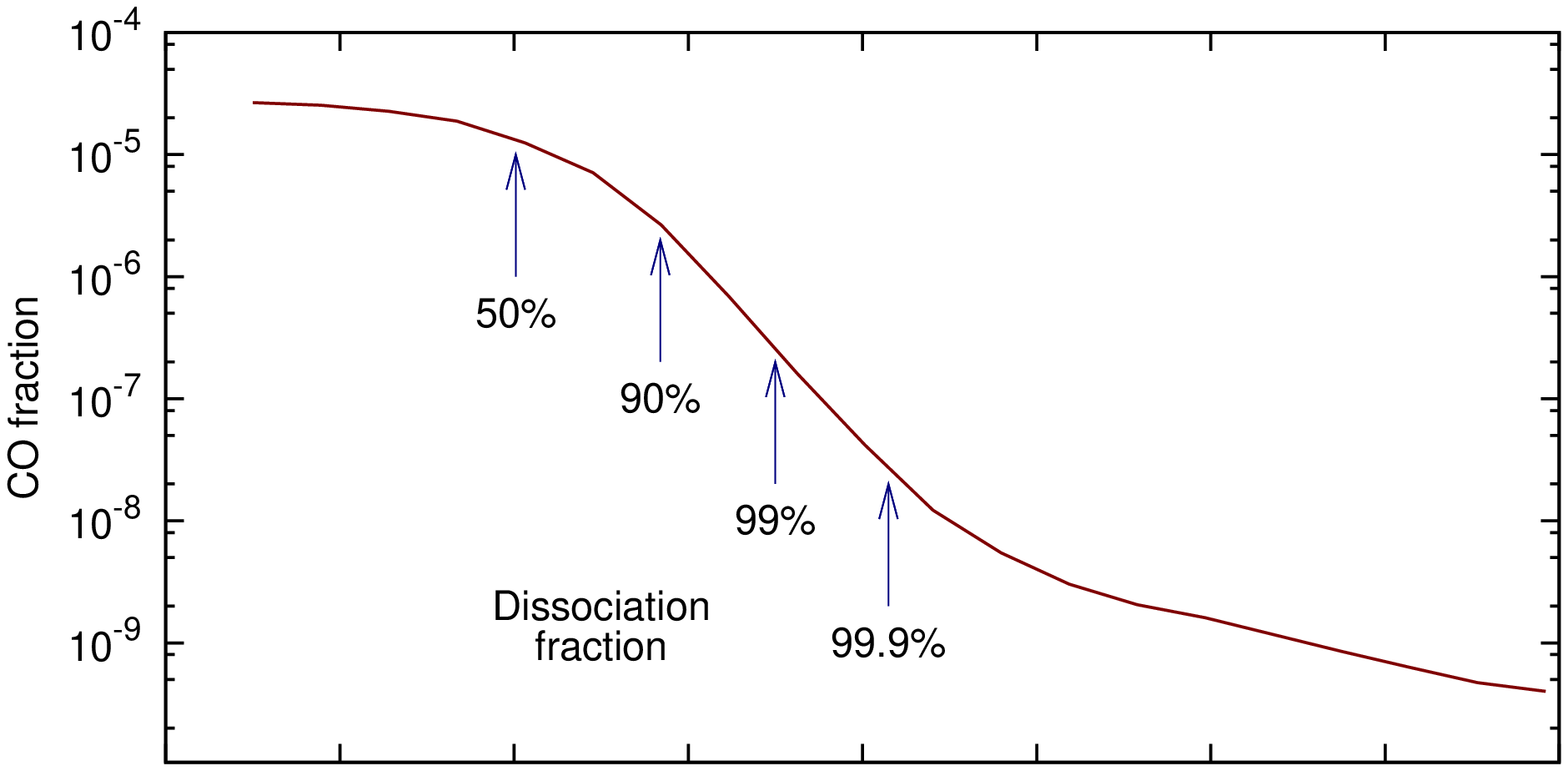}}
\centerline{\includegraphics[height=0.47\textwidth,angle=-90]{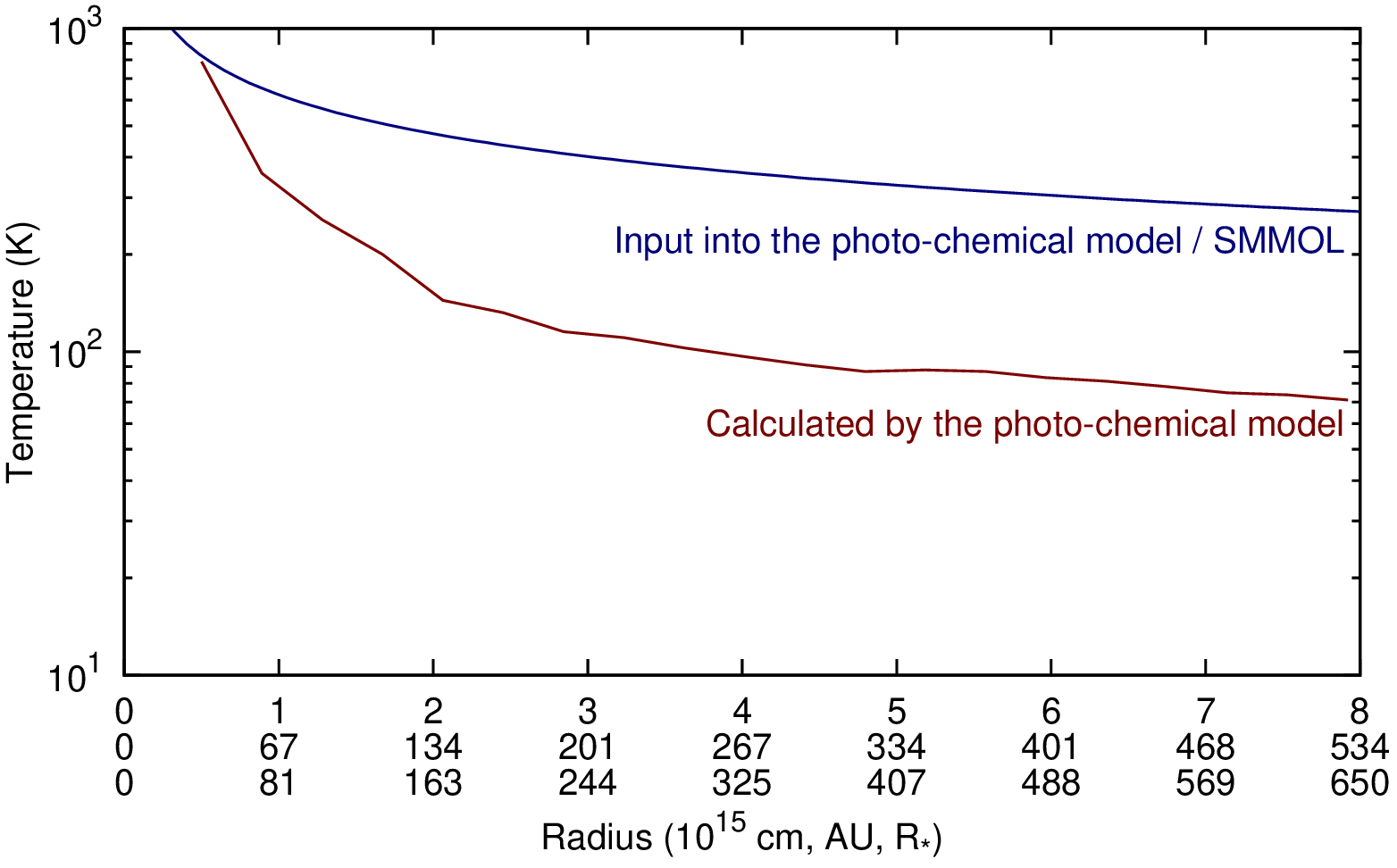}}
\caption{Radial properties of the toy photo-chemistry model. This model does not include heating by the central star, hence the difference between the temperature profiles in the bottom panel.}
\label{ArepoFig}
\end{figure}

To get a simple estimate of the abundance of CO that could be expected in a high-UV-field, metal-poor environment we set up the following toy model, which we base on 47 Tuc V1. We used the moving-mesh code {\sc arepo} \citep{Springel10}. Our version of the {\sc arepo} code has been modified to include time-dependent chemistry as presented by \citet{SGK14} (for full details see \citet{Glover07} and \citet{GC12}). The CO abundance is calculated assuming that the CO formation is limited by an initial radiative association step, and that the CO destruction rate is primarily through photodissociation. The gas is shielded from the ambient radiation field by dust and gas self-shielding, which we calculate using the approach of \citet{CGKB12}.

The physical model consists of an expanding shell of gas with a total mass of 9 $\times$ 10$^{-4}$ M$_\odot$, following a $M(r) \propto r^{-2}$ power law, starting from the assumed stellar surface at $R_\ast$ = 1.3 $\times$ 10$^{13}$ cm = 0.87 AU. The entire envelope corresponds to the mass ejected over 3000 yr\footnote{This value was chosen to be sufficiently larger than the typical envelope filling time of similar stars under solar-neighbourhood conditions ($\sim$1300--2200 yr; Section \ref{COExpectSect}). The exact radius and time are not critical to the results.} at a mass-loss rate of 3 $\times$ 10$^{-7}$ M$_\odot$ yr$^{-1}$. The gas initially follows a $T(r)\propto r^{-0.4}$ power law (e.g.\ \citealt{Groenewegen12b}) with the inner surface corresponding to the temperature of the stellar photosphere, taken to be 3517 K.

In this oxygen-rich environment, the fractional abundance of CO is determined by the carbon abundance: [C/Fe] = --0.13 $\pm$ 0.20 dex \citep{RCGS14}. The carbon abundance varies between [C/Fe] $\sim$ --0.5 dex for a few CNO-processed stars to [C/Fe] = 0.0 dex for the majority of the stellar population \citep{BHSG04}. We use [Fe/H] = $-0.72$ dex, following \citet{Harris10}, and [C/H] = --0.72 dex to conservatively model the carbon-rich end of the population. We assume a dust-to-gas ratio of 1:1000 \citep{MBvLZ11}, and [O/H] = --0.57 dex \citep{RCGS14}. The envelope is initially fully molecular, with all its carbon in the form of CO.

The CO abundance relative to protons was tracked just inside the inner boundary of the expanding shell, representing an expanding region of gas in a dynamic envelope. The code assumes an external UV field with a spectrum equivalent to that seen in the Solar Neighbourhood \citep{Draine78}, but with a flux that is 50$\times$ higher (i.e.\ 50 Habing units\footnote{A Habing unit is the integrated flux between 912 and 1110 \AA\ in the Solar Neighbourhood, i.e.\ $1.6 \times 10^{-3}$ erg s$^{-1}$ cm$^{-2}$}; see Section \ref{CODissSect} and Table \ref{ISRFTable}).

This model includes several important simplifications, which generally lead to an \emph{overestimate} of the CO abundance, and an overestimate of the CO dissociation radius. The most important of these assumptions are:
\begin{itemize}
\item The adoption of an ISRF from the Solar Neighbourhood, scaled upwards by a factor of 50 to the expected \emph{time-averaged} flux of the ISRF in 47 Tuc at 1076 \AA. The UV field in globular clusters also contains harder UV photons, which will lead to more rapidly dissociated CO, and is likely to result in significant quantities of CO$^+$, or even CO$^{++}$.
\item A lack of radiative heating from the central star, and a lack of thermal heating by circumstellar dust (collisional heating is included). This means the temperature of the CO envelope is colder than we expect (Figure \ref{ArepoFig}), protecting the CO in lower ro-vibrational states. Figure 2 of \citet{MGH88} shows that halving the CO excitation temperature has a small but noticeable effect on CO shell size.
\end{itemize}

Figure \ref{ArepoFig} shows the abundance of CO relative to protons as a function of dynamic radius. At the stellar surface, the carbon is entirely bound in molecules, with a CO:proton ratio of 2.67 $\times$ 10$^{-5}$. The abundance drops rapidly with distance due to photodissociation by the ambient UV field. By a radius of 2.8 $\times$ 10$^{15}$ cm (0.00092 pc, or 218 R$_\ast$), 90 per cent of the CO has been dissociated. This represents a shell $\sim$22$\times$ smaller than predicted by \citet{MGH88} for a similar Galactic star. Bearing in mind the simplifications above, we can expect the shell size to be smaller still, suggesting a close to inverse relation between ISRF strength and CO shell size.

\subsection{Predicting the CO line fluxes}

\begin{table}
 \centering
  \caption{Parameters used in the CO line flux models}
\label{ModelTable}
  \begin{tabular}{l@{\ }c@{\ }c@{\ }c}
  \hline\hline
   \multicolumn{1}{c}{Parameter} & \multicolumn{1}{c}{\clap{Symbol}} & \multicolumn{1}{c}{Unit} & \multicolumn{1}{c}{Value}\\
 \hline
\multicolumn{4}{c}{\it Used in both the {\sc arepo}+photochemical model and {\sc smmol}}\\
Gas temperature$^{[1]}$	& ...		& ...		& $T_\ast\!\left(\frac{R}{{\rm R}_\ast}\right)^{-0.4}$ \\
Dust:gas ratio$^{[2,3]}$& ...		& ...		&1:1000	\\
Initial CO/H$_2$	& ...		& ...		& $5.33 \times 10^{-5}$	\\
\hline
\multicolumn{4}{c}{\it Used in the {\sc arepo}+photochemical model}\\
Outflow velocity	& ...		& km s$^{-1}$	& 10 	\\
Wind density at $R_\ast$& ...		& cm$^{-3}$	& $2.63 \times 10^9$ \\
Wind density law	& ...		& cm$^{-3}$	& $n_0 \left(\frac{R}{{\rm R}_\ast}\right)^{-2}$ \\
Mass-loss rate		& ...	& M$_\odot$ yr$^{-1}$	& $3 \times 10^{-7}$ \\
Dust condensation	& $R_{\rm cond}$& cm		& $1.23 \times 10^{13}$	\\
Dust temp.\ at $R_{\rm cond}$ & ...& K		& [4]	\\
ISRF			& ...		& Habing	& 50	\\
\hline
\multicolumn{4}{c}{\it Used in {\sc smmol}}\\
Stellar radius$^{[2]}$	& $R_\ast$	& cm	 	& $1.23 \times 10^{13}$	\\
Stellar spectrum	& ...	& ...	 		& [2]	\\
Stellar mass		& $M_\ast$	& M$_\odot$	& 0.55 \\
Stellar luminosity	& $L_\ast$	& L$_\odot$	& 4824 \\
Stellar $T_{\rm eff}$	& $T_\ast$	& K	 	& 3623 \\
Dust condensation$^{[2]}$& $R_{\rm cond}$& cm		& $2.46 \times 10^{13}$	\\
Dust temp.\ at $R_{\rm cond}$ & ...& K	&	850$^{[2]}$  \\
Dust composition	& ...		& ...		& [2]	\\
 \hline
\multicolumn{4}{p{0.47\textwidth}}{Notes: [1] The photo-chemical model does not include the central star so the gas cools radiatively. Heat exchange with dust is implemented, but dust and gas heating by the central star are not, so the ejecta cools radiatively. {\sc Smmol} resets the temperature structure to this assumed power law. [2] Following \citet{MBvLZ11}. [3] Based on scaling a solar value of 1:200 by [Fe/H] = --0.72 dex. The ratio is limited by condensable metals to $<$1:778 \protect\citep{MBvLZ11}. [4] As parameterised by kinetic temperature law.}\\
 \hline
\end{tabular}
\end{table}

Millimetre/sub-mm CO lines at these mass-loss rates are optically thin (e.g.\ \citealt{RSOL08}). Given the $r^{-2}$ density law of our wind, we can expect the CO line flux to scale with the shell size, such that the CO line flux should be $\sim$22$\times$ lower than predicted by \citet{RSOL08} for a star losing mass at 3 $\times$ 10$^{-7}$ M$_\odot$ yr$^{-1}$. Other factors are important, notably the temperature structure of the atmosphere, which controls the rotational level populations of CO. We have therefore predicted line strengths from our envelope using the {\sc smmol} radiative transfer code \citep{RY01}. Table \ref{ModelTable} lists the parameters used in the model.

\subsubsection{SMMOL}

{\sc Smmol} is a spherical-geometry, non-LTE radiative transfer code which solves level populations with the accelerated lambda iteration (ALI) scheme, described by \citet{RH91} and \citet{SC85}. It relies on a discrete spatial grid and ray tracing for radiation transfer and is optimised for high-optical-depth regimes. 

{\sc Smmol} adopts dust properties from \citet{MBvLZ11}, which uses the {\sc Dusty} radiative transfer code \citep{NIE99} to model the spectral energy distribution of V1. This model uses a grain mixture of 88 per cent metallic iron \citep{OBA+88} and 12 per cent silicates \citep{DL84}, surrounding a star modelled by a {\sc bt-settl} model atmosphere spectrum \citep{AGL+03}. The gas temperature structure was taken from our photo-chemical modelling results. An accelerating wind was used, with a velocity at the inner region of 8 km\,s$^{-1}$, and a terminal velocity of 11 km\,s$^{-1}$, to mimic radiative acceleration of dust. This lowers the optical depth of the profile but, as our stars have low mass-loss rates, the resulting line flux differs negligibly from a constant 10 km s$^{-1}$ wind. A turbulent velocity of 2\,km\,s$^{-1}$ was adopted. Table \ref{ModelTable} lists a full set of physical parameters.

{\sc Smmol} modelled the observation as observed by a single 250-metre telescope with a perfect Gaussian beam. The CO envelope (0$\farcs$05) remains a point source at this resolution (1$\farcs$07), such that the conversion from antenna temperature to flux units (Janskys) represents the flux from the entire source as observed by ALMA.

\subsubsection{Results}

Table \ref{ModelResultsTable} lists the resulting line fluxes from {\sc smmol}. {\sc Smmol} clearly predicts line strengths below the upper limits of our observations, showing that photo-dissociation explains well our non-detection of the CO emission from these stars.

We stress that these fluxes are only order-of-magnitude estimates, as many wind parameters are not accurately known. The line flux is strongly affected by the temperature structure of the inner stellar wind. We have adopted a simple power law for our model, but the physical temperature will be significantly affected by the dust properties. The dust properties of our target stars are not well known, particularly the grain mineralogy, grain size and dust-production rate, though they are likely very different from AGB stars in the Solar Neighbourhood \citep{MBvLZ11}. This uncertainty therefore translates into a considerable uncertainty on the CO line strengths. When combined with the time-variability of the ISRF (Section \ref{ISRFSect}), which can exceed an order of magnitude, it becomes clear that we are limited to stating that the line fluxes will scale very approximately with ISRF strength, and that we would typically expect them to be below the sensitivity limit of our observations.

\begin{table}
 \centering
  \caption{Predicted line fluxes from the photo-chemical model of V1. Note these are conservatively bright estimates.}
\label{ModelResultsTable}
  \begin{tabular}{lcc}
  \hline\hline
   \multicolumn{1}{l}{Line} & \multicolumn{1}{c}{Observational} & \multicolumn{1}{c}{Predicted line fluxes}\\
   \multicolumn{1}{l}{\ } & \multicolumn{1}{c}{limit} & \multicolumn{1}{c}{from \sc smmol}\\
   \multicolumn{1}{l}{\ } & \multicolumn{1}{c}{(mJy\,km\,s$^{-1}$)} & \multicolumn{1}{c}{(mJy\,km\,s$^{-1}$)}\\
\hline
$J$ = 4$\rightarrow$3 & \nodata & 16 \\
$J$ = 3$\rightarrow$2 & $\lesssim$2000$^{[1]}$ & 6.1 \\
$J$ = 2$\rightarrow$1 & $<$183 $\pm$ 85$^{[2]}$ & 2.6 \\
$J$ = 1$\rightarrow$0 & \nodata & 0.1 \\
\hline
\multicolumn{3}{p{0.45\textwidth}}{$^1$APEX, $^2$ALMA. See Section \ref{ObsResultsSect}.}\\
\hline
\end{tabular}
\end{table}



\section{Discussion \& Conclusions}
\label{DiscSect}

In this paper, we have presented new ALMA observations of the four brightest stars in the globular cluster 47 Tucanae, examining the CO $J = 2\rightarrow1$ transitions. No source was detected. Having modelled the interaction between the ISRF and the circumstellar enviroment, we expect that the CO line fluxes from the giant stars in 47 Tuc are around two orders of magnitude below our observation limit.

For any given globular cluster giant, if the wind is optically thin, then the radius of CO dissociation should exhibit an approximately linear relation with metallicity and an approximately inverse relation with ISRF strength (Section \ref{CODissSect}). The radiation field should be highly stochastic, driven by the temporally variable population of hot white dwarfs. The population of young white dwarfs is not well determined in most clusters, therefore it is not immediately possible to identify which clusters will be the best candidates for observation. However, the ISRF can be expected to vary by several orders of magnitude within a cluster, both temporally and spatially (Table \ref{ISRFTable}). Targets far from the cluster centre may experience radiation fields that are weaker than those in the Solar Neighbourhood (though they will still contain harder radiation). Giant stars at large distances from their host clusters' centres may therefore provide the best metal-poor stars around which to detect stellar winds and measure their expansion velocities. Table \ref{ISRFTable} indicates that V2 and V3 may occasionally experience ISRFs that are weaker than the Solar Neighbourhood. While this seems not to be the case at present, we can hope that a similar situation can be found in other Milky Way globular clusters.

Since the higher stellar density forces the CO envelopes to be smaller than their Galactic counterparts, the CO(3--2) line (or higher rotational states) may be considerably brighter than CO(2--1), as these are better populated in the warmer regions close to the star (Table \ref{ModelTable}). We therefore suggest observations to detect CO line strengths and widths from globular cluster stars focus on:
\begin{enumerate}
\item stars at large radii from their host clusters and
\item high CO rotational states emitted closer to the star.
\end{enumerate}

We also advocate further observations of the circumstellar dust. Unexpectedly large amounts of dust are seen around many globular cluster stars \citep{SMM+10,MSZ+10,MBvLZ11}. Their spectra have strong silicate features \citep{LPH+06,vLMO+06,MBvLZ11}, but also an underlying continuum, possibly explained by metallic iron \citep{MSZ+10}. This contrasts with the notable absence of interstellar dust within clusters \citep{BMvL+08,MMN+08,MvLD+09,BMvL+09,BBW+09}, and suggests dust is being destroyed while still in the circumstellar environment. The highest-energy photons ($\sim$40 eV, $\sim$300 \AA) are likely to penetrate into the dust-forming regions. While they are of insufficient number to dissociate a large fraction of CO in this region, they may still be important in the photo-chemical formation of dust.

Finally, we note that this situation is unlikely to be limited to globular clusters. The dissociation of CO around AGB stars by nearby post-AGB stars and white dwarfs is largely a function of stellar density (although stellar age, elemental composition and the density of the surrounding ISM are also important). This source of CO dissociation may also become important in old open clusters, nuclear star clusters, and gas-poor galaxies. The implications for the formation, chemistry, evolution and survival of dust in such environments are significant.


\section*{Acknowledgements}
This paper makes use of the following ALMA data: ADS/JAO.ALMA\#2012.1.00333.S. ALMA is a partnership of ESO (representing its member states), NSF (USA) and NINS (Japan), together with NRC (Canada) and NSC and ASIAA (Taiwan), in cooperation with the Republic of Chile. The Joint ALMA Observatory is operated by ESO, AUI/NRAO and NAOJ.

This publication is based on data acquired with the Atacama Pathfinder Experiment (APEX). APEX is a collaboration between the Max-Planck-Institut fur Radioastronomie, the European Southern Observatory, and the Onsala Space Observatory. The project identifier for these data is O-092.F-9327(A).

G.S.\ was supported by the NSF (Award 1108645, An ALMA Reconnaissance of Distant Dying Stars).

We are grateful to the anonymous referee for their constructive feedback on the manuscript.

\appendix
\section{Other processes affecting the stellar envelope}
\label{StellarCOSect}

In the main text, we have shown that photo-dissociation of CO by interstellar UV radiation is a significant mechanism which reduces the CO envelope size. In this Appendix, we consider whether alternative processes acting on globular cluster stars may also affect the size of the CO envelope.

\subsection{Envelope size, filling time and resolution}
\label{COExpectSect}


In determining the effect of other processes, the relevant size-scales and time-scales must be compared to the extent and filling time expected for the CO envelope. In this section, we determine what those would be for an AGB star with the properties of those in 47 Tuc but with a Galactic ISRF incident upon it. We also examine the possibility that the CO might be resolved by the ALMA interferometer.

The limiting size of a CO envelope in a Galactic environment is usually set by its dissociation by the ISRF. \citet[][their figure 3]{MGH88} predict that 90 per cent of the CO in the envelope of a Galactic AGB star with a mass-loss rate of a few $\times$ 10$^{-7}$ M$_\odot$ yr$^{-1}$ will be dissociated by a radius of $\sim$0.013--0.023 pc. This equates to $\sim$0.6$^{\prime\prime}$--1.1$^{\prime\prime}$ at the distance of 47 Tuc. Assuming a typical wind velocity for less-evolved stars of 10 km s$^{-1}$ \citep{WlBJ+03,Groenewegen14}, a typical CO envelope would therefore be filled within 1300 to 2200 years. If CO is more rapidly dissociated by interstellar UV, we can expect the radius to be smaller and the filling time to be correspondingly shorter.

This would preclude a much larger envelope, which could be resolved by the ALMA beam. Since the winds are optically thin, the dissociation radius is mainly set by the time taken for a CO molecule to absorb a UV photon. A faster wind may therefore lead to a larger envelope, but the velocity and envelope size will not be directly proportional. Larger envelopes will be less self-shielding, as the optical depth in the dissociating lines will decrease. It is unlikely that an envelope could reach the velocity needed to extend beyond the 2$^{\prime\prime}$ ALMA beam ($\gtrsim$20 km s$^{-1}$), and virtually impossible for it to reach the $\gtrsim$60 km s$^{-1}$ needed to be diluted by a sufficient factor that it becomes unobservable in the central beam. This would still be detectable by the larger $15.7^{\prime\prime} \times 15.7^{\prime\prime}$ box, so we consider it very unlikely that the large-scale structure of our envelopes are resolved by the ALMA interferometer.

\subsection{Episodic or variable mass loss}
\label{COVarSect}

If mass loss is episodic or variable, it is feasible that we could see strong mid-IR emission from recent dust production near the star, while the CO envelope further from the star traces a period of lower mass-loss rate. Given the chromospheric and CO mass-loss rates are discrepant by a factor of $\sim$10 in each case, any variability or episodic nature must be correspondingly large, i.e.\ a mass-loss rate variation of 10$\times$ or, alternatively, mass loss occurring only 10 per cent of the time. While the chances of finding any four stars going through this state is high, the chances of finding four particular stars (in this case the most luminous) simultaneously going through an episode of enhanced mass loss is therefore $\sim$10$^4$:1 against.

This probability only holds if the enhancement in dust production is uncorrelated with stellar luminosity. Such a correlation occurs, for example, during the bright phase of a thermal pulse. \citet{GWG+10} model the thermally-pulsating (TP) phase of a globular-cluster-like AGB star to be $\sim$1.2--1.8 Myr in length. If a star leaves the AGB every 80\,000 years \citep{MBvL+11}, this lifetime equates to 15--23 stars on the TP-AGB, four of which we observe, the remainder of which will be at luminosities at or below the RGB tip. The bright phase of a thermal pulse is relatively short compared to the pulse cycle: $\sim$0.3 per cent \citep{Herwig05}. The chances of a particular star being in the bright phase of a thermal pulse are therefore $\sim$0.3 per cent, and the chances for four such stars are $\sim$1 in 12 billion.

\subsection{Self-destruction by chromospheric heating and/or stellar UV}
\label{COStarSect}

Although CO bands are present in the near-IR spectra of these stars \citep{LWH+05}, it is possible that they could produce enough UV photons to destroy their own CO envelopes. Metal-poor stars are warmer than their solar-metallicity counterparts due to reduced metal-line opacity in their atmospheres. Escaping stellar UV could lead to the destruction of CO near the stellar surface. A {\sc BT-Settl} model atmosphere \citep{AGL+03} at 4000 L$_\odot$ and 3300 K emits $3 \times 10^{25}$ photons per second below the 1120-\AA\ CO photo-dissociation limit. By comparison, a mass-loss rate of $3.4 \times 10^{-7}$ M$_\odot$ yr$^{-1}$ and efficient condensation (i.e.\ a CO:H$_2$ mass ratio of 1:500) yields $3 \times 10^{41}$ CO molecules produced per second. Many more CO molecules are produced than there are UV photons to dissociate them.

While the {\sc BT-Settl} atmospheres do not model chromospheric or shock heating in the outer atmosphere, the heating cannot compensate for the factor of 10$^{16}$ between the modelled number of ionising photons and number required to dissociate all the CO the star produces. Assuming a typical wavelength of 1105 \AA, $3 \times 10^{41}$ photons provide an energy of $8 \times 10^{20}$ W, or $10^{-14}$ W m$^{-2}$ as observed from Earth. Escaping photons would therefore be easily detectable in the \emph{GALEX} FUV filter for all realistic spectral energy distributions. None of the four sources is present in catalogues from the satellite\footnote{{\tt http://galex.stsci.edu/}}, which is complete over 47 Tuc to a far-UV AB magnitude of $\approx$20.0 ($\approx$1 $\times$ 10$^{-16}$ W m$^{-2}$), approximately two orders of magnitude more sensitive than would be necessary to detect such a UV field.

Reduced metal-line and dust cooling in pulsation-driven shocks presumably mean that the post-shocked gas further out in the atmosphere also cools more slowly, leading to more effective dissociation of CO further from the stellar surface. However, observations show the expected column depths of CO absorption lines in the $H$-, $K_{\rm s}$- and (archival) $L$-band VLT/CRIRES data \citep{LWH+05}, indicating that CO must survive to at least a few stellar radii. The observed CO absorption and lack of observed UV emission from these stars means that self-destruction of CO by the stars that produce it appears unlikely.

\subsection{Stellar encounters}
\label{COStarsSect}

Stellar encounters are common in globular clusters, and are thought to be responsible for a lack of planets and binary stars, and a prevalence of stellar exotica \citep{GBG+00,DS01,WSBF05,KDMA+08,SGHL09}. While stellar encounter rates have previously been computed for 47 Tuc, published values are not specifically applicable to giant stars' envelopes (e.g.\ \citealt{DB95}).

The encounter timescale ($\tau_{\rm enc}$) between an AGB star and any other star in the cluster can be computed as:
\begin{equation}
\tau_{\rm enc} \approx 7 \times 10^{10}\ {\rm yr} \frac{10^5\ {\rm pc}^{-3}}{n}\frac{v_{\rm enc}}{10\ {\rm km s}^{-1}}\frac{{\rm R}_\odot}{R_{\rm enc}}\frac{{\rm M}_\odot}{<\!M_\ast\!>} ,
\end{equation}
where $n$ is the stellar number density, $v_{\rm enc}$ is the typical relative velocity at the start of the encounter, $R_{\rm enc}$ is the maximum distance at which an encounter is deemed to occur and $<\!M_\ast\!>$ is the typical stellar mass (e.g.\ \citealt{DS01}). We conservatively assume an envelope radius of 0.03 pc (Section \ref{COExpectSect}), hence conservatively adopt $R_{\rm enc} = 0.06$ pc. We assume $v_{\rm enc}$ = 22 km s$^{-1}$, or twice the radial component of the central velocity dispersion of the cluster \citep{Harris10}. Given an average stellar density within the half-light radius of 1720 M$_\odot$ pc$^{-3}$ and d$N$/d$m$ $\propto$ $m^{-1.2}$ \citep{MK10}, bounded by a planetary mass (0.08 M$_\odot$) and the maximum sub-giant mass ($\sim$0.8 M$_\odot$), we obtain $<\!M_\ast\!>$ $\gtrsim$ 0.2 M$_\odot$ and $n$ $\lesssim$ 8600 pc$^{-3}$.

We conservatively derive $\tau_{\rm enc}$ $\gtrsim$ 340\,000 years, though the possible range of values allow for $\tau_{\rm enc} > 1$ Myr. This is much longer than the 2300-year envelope filling timescale. While a fraction of stars may undergo envelope-disrupting stellar encounters at some point in their dust-producing phase, this is extremely unlikely to affect all four observed stars simultaneously.


\subsection{Disruption in the cluster potential}
\label{COPotSect}

A circumstellar envelope will experience tidal disruption or `spaghettification' as it interacts with the cluster potential, which becomes significant once it experiences significant differential acceleration over the filling timescale (which is considerably shorter than the $\sim$1 Myr orbital period). We can assume a 0.06-pc wide envelope in a radial orbit in a \citet{Plummer1911} potential of characteristic radius 7.8 pc \citep{LKL+10}, and calculate the change in distance between opposite ends of the envelope. We find a maximum differential acceleration of $2 \times 10^{-11}$ m s$^{-2}$, corresponding to a tidal elongation of one part in 10$^4$. Tidal deformation of the envelope is therefore negligible.



\subsection{Sweeping by hot Halo gas}
\label{COBowSect}

There is a general paucity of interstellar media in globular clusters (e.g.\ \citealt{SWFW90,vLSEM06,vLSP+09}), with interstellar media detected in only M15 and 47 Tuc \citep{FKL+01,ESvL+03,BWvL+06,vLSEM06}. This strongly suggests that globular clusters are efficiently cleared of their ICM by some process, on timescales of $\lesssim$1 Myr \citep{BMvL+08}. We model that this is because it is ionised by hot white dwarfs and post-AGB stars, before being later cleared on slightly longer timescales (a few Myr; \citep{MZ15a}). However, this model does not include ablation by the passage of the cluster through the hot gas of the surrounding Galactic Halo, which is another leading contender \citep{FS91}.

The stand-off distance of a bow shock ($R_0$), i.e.\ the minimum distance from the bow shock to an isolated, mass-losing star, depends on four parameters: the mass-loss rate of the star and the expansion velocity of its wind ($\dot{M}$, $v_{\rm exp}$), the interstellar medium density ($\rho_{\rm ISM}$) and the relative velocity of the star with respect to the Halo gas ($v_\ast$). $R_0$ is given by (\citet{CKvM+12}, and references therein\footnote{Note that this momentum balance assumes cool Halo gas. Balancing ram pressure against gas pressure gives values of $R_0$ about three times larger.}):
\begin{equation}
R_0 = \sqrt{\frac{\dot{M} v_{\rm exp}}{4 \pi \rho_{\rm ISM} v_\ast^2}} .
\end{equation}

Depending on the rotational coupling between the Halo and Milky Way, $v_\ast$ lies between 57 km s$^{-1}$ (co-rotating Halo) and 190 km s$^{-1}$ (non-rotating Halo) \citep{KG95}. The internal velocity dispersion within 47 Tuc is negligible in comparison (11 km s$^{-1}$; \citealt{Harris10}). No good determinations exist for $\rho_{\rm ISM}$; estimates at 3.1 kpc from the Galactic Plane are typically $n_{\rm H} \sim 0.01$ cm$^{-3}$, but vary by an order of magnitude \citep{ABM+93}. \citet{TC93} estimate an electron density of $n_{\rm e} \approx 0.007$ cm$^{-3}$ at this location (we can expect $n_{\rm e} \approx n_{\rm H}$ in the ionised Halo gas), and the more-recent spherical model by \citet{MB13} predicts a lower value of $n_{\rm e} \approx 0.0002$--0.002 cm$^{-3}$.

Assuming $\dot{M}$ $\approx$ $10^{-7}-10^{-6}$ M$_\odot$ yr$^{-1}$, $v_{\rm exp}$ = 4--15 km s$^{-1}$ and $n_{\rm H} \sim 10^{-4}-10^{-1}$ cm$^{-3}$, $R_0$ = 0.02--38 pc, always greater than the expected CO shell radius. Additionally, these stars are not isolated, but are surrounded by a hot, ionised ICM \citep{MZ15a}, which should provide additional shielding out to $R_0 \approx 0.3$--98 pc. We conclude that ram-pressure stripping by Halo gas should not be effective at removing circumstellar CO envelopes.





\label{lastpage}

\end{document}